\documentclass[11pt,a4paper]{article}
    \usepackage{jheppub}
    \usepackage{caption,subcaption}
    \usepackage{tikz,tipa}
    \usetikzlibrary{arrows,shapes}

\newcommand{\g}{\gamma}

\begin{document}

\title{A new energy bound for Einstein-Scalar theory in AlAdS$_4$ and holographic bound for deformed CFT$_3$}

\author{Krai Cheamsawat}
\emailAdd{krai.cheamsawat15@imperial.ac.uk}

\affiliation{Theoretical Physics Group, Blackett Laboratory, Imperial College, London SW7 2AZ, United Kingdom}


\abstract{In this work, we derive an upper bound on energetic quantities, namely vacuum energy and free energy, for static solutions of Einstein-Scalar theory in four dimensional asymptotically locally Anti-de Sitter(AlAdS) spacetime with a nontrivial scalar potential where the scalar field mass parameter($m^2$) is equal to 0 or -2. This system is the holographic dual of strongly coupled conformal field theory(CFT) in three dimensions being deformed by a relevant or marginal scalar operator of conformal dimension $\Delta=1, 2, 3$. The bound is derived from a purely gravitational perspective regardless of the inhomogeneity of the static conformal boundary of AlAdS and the source of the deformation. We demonstrate the bound in simple settings and check the consistency with the known previous bounds.}

 \keywords{AdS/CFT correspondence, Conformal Field Theory, Black Holes }

\maketitle


\section{Introduction}
\paragraph{}The AdS/CFT correspondence\cite{Maldacena:1997re}\cite{Aharony:1999ti}\cite{Witten:1998qj} or informally known as holographic duality has proved to be greatly useful in the vast area of theoretical physics both directly related to its original motivation such as string theory, quantum gravity, black hole physics and its applications to neighbouring fields such as quantum chromodynamics dynamics(QCD), condensed matter physics, quantum information theory, etc. In this paper, we consider the scenario of gravity in asymptotically locally Anti-de Sitter (AlAdS) spacetime\footnote{Precise definition and the properties of AlAdS spacetime can be found in \cite{marolf2014conserved}.} which is dual to a holographic quantum field theory in curved spacetime(a review in this topic can be found in \cite{Marolf:2013ioa}) with one extra ingredient other than pure gravity; namely a scalar field is added to the bulk gravity theory in AlAdS. This additional scalar field in the bulk is dual to a deformation of the dual conformal field theory where the dimension of the dual operator depends on the mass of the scalar field. In the limit where the scalar field, $\phi$, is allowed to backreact on the geometry, the details of the bulk geometry will be strongly affected by the details of scalar field such as its boundary conditions and potential $V(\phi)$. This topic has received attentions and various important results have been found, for example, \cite{DeWolfe:1999cp}\cite{Gubser:2000nd}\cite{Bianchi:2001kw}\cite{DHoker:2002nbb}\cite{Henneaux:2006hk}. Yet still, in AlAdS$_4$, the structure of the theory is distinctively simple enough so that some general and very nontrivial results can be drawn without too many restrictions on its structure such as isometries of the spacetime and the type of potentials.

\paragraph{}The main question to be answered in this paper is: \textit{How do the properties of boundary conditions and potentials of the scalar field affect or constrain the properties of static solution in Einstein-Scalar theory in AlAdS$_4$?} More specifically, how do these affect the physical quantities of solutions? And on what condition do they dictate the existence of those constraints on physical quantities? In this regard, for pure gravity in AlAdS$_4$, the answer has been given in terms of the free energy bound(which becomes the vacuum energy bound in the zero temperature limit) in \cite{Hickling:2015tza}. The similar results for gravity with the massless scalar field without a potential in AlAdS$_4$ are shown in \cite{Fischetti:2016vfq}. 

\paragraph{}By the nature of Einstein-Scalar theory, the choices of potentials for the scalar field is infinite, therefore we keep our discussion as general as possible by not assuming a specific form of the potential in our derivation.  As the results whose details will be laid out later in this paper show, an upper bound on physical quantities of static solutions of Einstein-Scalar theory can be given for scalar fields whose potential satisfies a specific condition. When the condition is met, the upper bound is guaranteed to exist. The main result can be stated briefly as the following:

\paragraph{Main result} For static solutions of Einstein-Scalar theory in AlAdS$_4$ with conformal boundary topology $\mathbb{R} \times \Sigma$ where $\Sigma$ is a static arbitrary curved two dimensional space\footnote{$\Sigma$ should have all the usual properties such as being smooth and nonsingular.} with finite volume and no singularity or event horizon on the conformal boundary is allowed. The scalar field $\phi$ is subjected to boundary conditions determined by two functions $\alpha(x),\beta(x)$ which are coefficients of the near boundary expansion in the radial direction of the bulk where $x$ is the spatial coordinates on the conformal boundary. Specifically, in this paper, we only consider the scalar field which corresponds to the operator of dimension $\Delta=1,2,3$. The condition on the potential of the scalar field is set by the function $B(\phi)=V(\phi)/\ell^2+V(\phi)^2/6-V'(\phi)^2/4$. When $B(\phi)\geq 0$ for all $\phi$, the following bound is guaranteed to exist:

\begin{itemize}
\item Finite temperature: Free energy for the finite temperature solution is bounded above
\begin{equation}
F \leq  4\pi TC_T\left( 2\pi \chi(\mathcal{H})-\frac{1}{4}\int_{\mathcal{H}}(\mathcal{D}\varphi)^2 \right)+C_Tf_{\text{bc}}(\Delta)\int_{\Sigma} \alpha(x)\beta(x)
\end{equation}

\item Zero temperature: as the $T\to 0$ limit is taken, the free energy bound becomes the vacuum energy bound
\begin{equation}
E_{\text{vac}} \leq C_Tf_{\text{bc}}(\Delta)\int_{\Sigma} \alpha(x)\beta(x)
\end{equation}

where $\mathcal{H}$ is a killing horizon whose surface gravity is related to its Hawking temperature by $\kappa=2\pi T$, $\chi(\mathcal{H})$ is the Euler characteristic of the horizon and $\mathcal{D}\varphi$ is the derivative of $\varphi$  which is the value of $\phi$ on the horizon.
\end{itemize}
In both of these cases, $f_{\text{bc}}$ is a coefficient that depends on the dimension of the dual operator $\Delta$ and boundary conditions of the scalar field and $C_T$ is an effective central charge\cite{Osborn:1993cr} of the dual CFT which is related to the bulk gravity quantities, AdS length $\ell$ and Newton constant $G_4$ by 
\begin{equation}
C_T = \frac{\ell^2}{16\pi G_4}
\end{equation}
which is taken to be large($C_T\gg1$) in order to make the bulk classical gravity plus scalar field system be the valid holographic dual of the dual CFT\cite{Aharony:1999ti}.

\paragraph{}This paper is organised as the following. The setup and the basic properties of Einstein-Scalar theory in AlAdS spacetime will be described in section(\ref{sec2}). The analysis of equations of motion, derivation of the general results and examples of potentials of the scalar field that obey the bound based on the condition given by function $B(\phi)$ is given in section(\ref{sec3}). In section(\ref{sec4}), the main result in terms of the upper bound on free energy and vacuum energy is given. An alternative perspective on the holographic bound is discussed and solutions to some simple cases are given in section(\ref{sec5}). Verification of the bound based on numerical black hole and soliton solutions for a scalar field with general boundary conditions is shown and extra results on the upper bound on the mass of solitons in designer gravity are demonstrated in section(\ref{sec6}). Conclusion and discussion of the results and some open problems based on these results are given in section(\ref{sec7}). Details of the near boundary and near horizon expansions for quantities in the bulk are given in the appendix.\ref{app}. The derivation of on-shell action is given in appendix.\ref{Sonshell}.

\paragraph{Notation convention} Indices for full spacetime coordinates are denoted by capital latin $A,B,\ldots \in \{0, 1,2, 3\}$. Indices for coordinates in spatial directions are denoted by lower case latin starting from $i,j,\ldots$ running from $1$ to $3$. The radial coordinate of bulk spacetime is usually denoted by $r$ and the context in which $r$ is used will be clarified at the beginning of each discussion throughout the paper. Greek indices refer to coordinates in timelike conformal boundaries of AlAdS spacetime, in this case the full spacetime coordinates will be split into $x^{A} = \{r, x^{\mu}\}$. Lowercase latin indices $a,b,\ldots$ denote the coordinates on two dimensional spacelike surfaces at conformal boundaries or horizons, such that $x^{\mu} = \{ t, x^{a}\}$.  

\section{Setup} \label{sec2}
In this section, we review some basic properties of Einstein-Scalar theory in AlAdS spacetime and introduce the spacetime frame namely, the "optical frame" which we use to describe general static solutions in Einstein-Scalar theory and derive the main result.  
\subsection{Generalities}
We consider gravity and a scalar field in four dimensions with the action for the metric $g_{AB}$ and scalar field $\phi$
\begin{equation}
S_{\text{total}} = -\frac{1}{2\kappa_4^2}\int_{\mathcal{M}} d^4x\sqrt{-g}\left(R(g)-\frac{1}{2}(\partial\phi)^2-V(\phi)\right)+S_{\text{bndy}}
\end{equation}
where $(\mathcal{M}, g_{AB})$ defines the bulk geometry and $S_{\text{bndy}}$ is the boundary term\footnote{To obtain a well-defined holographic dictionary and finite on-shell action for solutions in the theory one needs Gibbons-Hawking-York term($S_{GHY}$) and holographic counter term action($S_{\text{ct}}$) such that $S_{\text{bndy}}=S_{GHY}+S_{\text{ct}}$. We omit it here because it doesn't contribution to the equations of motion.} of the action which is being integrated on the timelike conformal boundary $\partial \mathcal{M}$ of AlAdS. Generically, the potential of the scalar field takes the form  $V(\phi) = 2\Lambda +\frac{1}{2}m^2\phi^2+\mathcal{O}(\phi^3)$ such that the AdS$_4$ spacetime with $\Lambda = -3/\ell^2$ is the solution when $\phi=0$, where $\ell$ is the AdS length and $2\kappa_4^2=16\pi G_4$. The equations of motion for the full spacetime metric  $g_{AB}$ and scalar field $\phi$ are
\begin{equation}\label{main}
\begin{aligned}
R_{AB} &= \frac{1}{2}\partial_{A}\phi\partial_{B}\phi+\frac{1}{2}g_{AB}V(\phi)   \\
\nabla^2\phi &= V'(\phi) 
\end{aligned}
\end{equation}
Since the scalar field backreacts on the geometry, even though the spacetime is asymptotically locally AdS, the geometry deep inside the bulk can be drastically different from pure AdS spacetime as a consequence of the potential for the scalar field and the profile of the asymptotic behaviour of the scalar field itself.

\paragraph{}Nevertheless, one universal feature of Einstein-Scalar theory is that the potential $V(\phi)$ determines the asymptotic behaviour of the scalar field near the conformal boundary. Let's discuss the general case of a $d+1$ dimensional bulk first. According to the holographic dictionary, adding the scalar field to the pure gravity sector in AlAdS$_{d+1}$ corresponds to adding the deformation by a single trace operator to the dual CFT$_{d}$ in the following manner\footnote{This is not always the case since the holographic CFT can also spontaneously obtain the vev, $\langle \mathcal{O} \rangle$ without the source of $\mathcal{O}$ turned on. The well-known example is a holographic superconductor\cite{Gubser:2008px}\cite{Hartnoll:2008kx}.}
\begin{equation}
S_{\text{CFT}} \to S_{\text{CFT}}+ \int d^dx\sqrt{\bar{g}}J(x)\mathcal{O}(x)
\end{equation}
where $J(x)$ is a source and $\mathcal{O}(x)$ is a spin zero, single trace operator of mass dimension $\Delta$ and $\bar{g}$ is the metric of spacetime on which the CFT$_{d}$ lives. From the analysis of the scalar field equation (\ref{main}) with $V'(\phi)=m^2\phi+\mathcal{O}(\phi^2)$ \cite{skenderis:2002wp} the mass parameter is related to the conformal dimension of its dual operator as $m^2\ell^2=\Delta(\Delta-d)$ where the two solutions of $\Delta$ for the given $m^2$ are the following
\begin{equation}
\begin{aligned}
\Delta_{\pm} = \frac{d}{2}\pm&\sqrt{\frac{d^2}{4}+m^2\ell^2} \\
\end{aligned}
\end{equation}
since $\Delta_{\pm}$ have to be always real, $m^2$ can be negative but cannot be less than the Breitenlohner-Freedman bound\cite{Breitenlohner:1982bm}\cite{Breitenlohner:1982jf} given by
\begin{equation}
m^2_{BF}\ell^2 = -\frac{d^2}{4}.
\end{equation}
Furthermore, there exists the lower bound on $\Delta$ called the "unitarity" bound $\Delta \geq \frac{d}{2}-1$. The parameter $m^2$ in the potential is fixed, $\Delta_{\pm}$ are too and subsequently the asymptotic behaviour of the scalar field near the boundary (where $r\to \infty$ in this radial coordinate) takes the form
\begin{equation}
\phi(r,x) = \alpha(x)\frac{\ell^{2\Delta_{-}}}{r^{\Delta_{-}}}+\ldots+\beta(x)\frac{\ell^{2\Delta_{+}}}{r^{\Delta_{+}}}+\ldots  \label{phiexpansion}
\end{equation}
where $\ldots$ are subleading terms whose coefficients depend on $\alpha, \beta$ \footnote{Where the coefficients of the terms in first ellipsis in (\ref{phiexpansion}) depend on $\alpha$ while the coefficients of terms in second ellipsis depend on $\alpha$ and $\beta$.}. Since an AlAdS spacetime is not globally hyperbolic, to have a well-defined dynamics of the theory one needs to specify the boundary condition for the metric and scalar field at every time slice. This fixing of boundary conditions for the scalar field amounts to fixing the boundary data $\alpha(x)$, $\beta(x)$ or the relation between them. For different values of $m^2$, allowable boundary conditions are different \cite{Henneaux:2006hk}\cite{Ishibashi:2004wx}. Since we only consider the AlAdS$_4$ spacetime, we then consider only the scalar field dual to the operator of conformal dimensions $\Delta=1,2,3$(the reason for this will be clarified later) hence the mass parameter of the scalar field are $m^2\ell^2=-2,0$. For these values of $m^2$, allowed boundary conditions for the scalar field are the following:

\paragraph{$\blacktriangleright m^2\ell^2=-2$}: there are three possible kinds of boundary conditions.

\begin{enumerate}
\item Dirichlet boundary condition : $\alpha(x)$ is identified as a source $J(x)$ while $\beta(x)$ must be solved from the equation of motions and it is identified as a one-point function $\langle \mathcal{O}(x)\rangle \sim \beta(x)$ with conformal dimension $\Delta = \Delta_{+}=2$. This is also called "\textit{Regular Quantisation}".

\item Neumann boundary condition : $\beta(x)$ is identified as a source $J(x)$ while $\alpha(x)$ must be solved from the equation of motions and it is identified as a one-point function $\langle \mathcal{O}(x)\rangle  \sim \alpha(x)$ with conformal dimension $\Delta = \Delta_{-}=1$. This is also called "\textit{Alternative Quantisation}".

\item General boundary condition\footnote{This is sometimes called "mixed boundary condition" or "Robin boundary condition".} : $\beta(x)$ is allowed to be local function of $\alpha(x)$ such that $\beta=\frac{\partial W(\alpha)}{\partial\alpha}$ or vice versa where $W(\alpha \text{ or } \beta)$ is an arbitrary function of $\alpha$ or $\beta$ (depending on whether it is generalised from a regular or alternative quantisation perspective \cite{Witten:2001ua}). The function $W$ corresponds to deforming the CFT$_{3}$ with a multitrace deformation\footnote{This $W$ is different from the "superpotential" in the supersymmetry context which is also normally denoted by $W$(and its other variant notations) as well.} \cite{Papadimitriou:2007sj}\cite{Vecchi:2010dd}\cite{Sever:2002fk}
\begin{equation}
S_{\text{CFT}} \to S_{\text{CFT}} + \int d^3x\sqrt{\bar{g}}W(\mathcal{O})
\end{equation}
particularly in asymptotically AdS$_{4}$ spacetime, this boundary conditions has been thoroughly studied in the context of "Designer gravity".\cite{Hertog:2004rz}\cite{Hertog:2004ns}\cite{Faulkner:2010fh}\cite{Amsel:2007im}\cite{Hertog:2004dr}.
\end{enumerate}

\paragraph{$\blacktriangleright m^2\ell^2=0$}: Only the dirichlet boundary condition can be applied where $\alpha(x)\sim J(x)$ and $\beta(x)$ is identified with a one-point function of the operator with conformal dimension $\Delta=3$. Note that even though the dimension of the dual operator is marginal, when the potential the scalar field has nontrivial higher order in $\phi$ such that $\ell^2V(\phi)=-6+\mathcal{O}(\phi^3)$ then the dual deformation of the CFT will be \emph{marginally relevant} or \emph{marginally irrelevant} depending on the higher order terms in $V(\phi)$.

Here we also note that, what is called "Dirichlet" or "Neumann" boundary condition in some literatures may refer to the boundary condition that $\alpha$ or $\beta$ are strictly fixed to be zero respectively. 

\subsection{Optical geometry}
\paragraph{}Additionally, we also restricted ourselves to consider only static AlAdS spacetime with conformal boundary, $\partial\mathcal{M}$. When the metric on the conformal boundary has no singularity or event horizon\footnote{Such a situation is, for example, CFT that lives on the black hole background. In which case the holographic duals are in black funnel or black droplet phases\cite{Hubeny:2009kz}\cite{Caldarelli:2011wa}\cite{Santos:2012he}} such that the $tt$-component of the metric always has a definite sign, the metric can be expressed in the ultrastatic frame where the three-dimensional spacetime has topology $\partial\mathcal{M} \simeq \mathbb{R}\times\Sigma$. $\Sigma$ is an arbitrary compact static two-dimensional manifold and its conformal boundary metric is 
\begin{equation}
d\bar{s}^2 = \bar{g}_{\mu\nu}dx^{\mu}dx^{\nu} =-dt^2+\bar{g}_{ab}(x)dx^adx^b =-dt^2+d\Sigma^2  \label{boundary}
\end{equation}
where $(\partial\mathcal{M}, \bar{g}_{\mu\nu})$ and $(\Sigma,\bar{g}_{ab})$ define the geometry of the conformal boundary and its spatial part respectively. With these conditions, we can express the generic metric of static AlAdS spacetime with such a boundary in terms of the "optical frame" metric:
\begin{equation}
ds^2 = g_{AB}dx^{A}dx^{B} = \frac{\ell^2}{Z(x)^2}\left(-dt^2+g_{ij}(x)dx^idx^j\right)  \label{optical}
\end{equation}
where the Riemannian manifold $(M,g_{ij})$ is called "optical geometry". Deep inside the bulk, it can end on a killing horizon(either extremal or non-extremal) or the union of them, which will be now called $\mathcal{H}$\footnote{A multi-component horizon such that $\mathcal{H}=\bigcup_{i=1}^{N}\mathcal{H}_i$ is allowed in our consideration as well.}. Strictly speaking, $\mathcal{H}$ is not the boundary of the optical geometry but rather an asymptotic end. Having this clarified, from now on, the boundary of the optical geometry will be referred to $\partial{M} \simeq \Sigma \cup \mathcal{H}$.


\section{Holographic bound} \label{sec3}
In this section, we show details of the derivation of the key relation in terms of a single elliptic PDE that will lead to the holographic bound which is the main result of this paper. Then we discuss some examples of potentials for the scalar field which satisfy the bound.
\subsection{Derivation \& General Result}
By writing the equations of motion (\ref{main}) in terms of the optical metric (\ref{optical}) and setting $\partial_{t}\phi,\partial_tZ,\partial_tg_{ij}=0$. The $tt, ij$-components of the Einstein equation and the scalar field equation now take the form

\begin{itemize}
\item $tt$-equation : 
\begin{equation}
\frac{1}{Z}D^2Z = \frac{\ell^2}{2Z^2}V(\phi) + 3\frac{(DZ)^2}{Z^2}
\end{equation}
\item $ij$-equation : 
\begin{equation}
R_{ij}+\frac{2}{Z}D_iD_jZ =\frac{1}{2} D_i\phi D_j\phi
\end{equation}
\item{$\phi$-equation} : 
\begin{equation}
D^2\phi -2\frac{D^iZ}{Z}D_i\phi = \frac{\ell^2}{Z^2}V'(\phi)  \label{phieqn}
\end{equation}
\end{itemize}
and by the $tt$-equation, there is an additional equation which is useful for computing the on-shell action,
\begin{equation}
D^2\left(\frac{1}{Z^2} \right) = -\frac{\ell^2}{Z^4}V(\phi)     \label{Veqn}
\end{equation}
where $D_i, R_{ij}$ are the covariant derivative and Ricci tensor with respect to the optical metric $g_{ij}$. The key relation can be derived by considering the following tensor of solutions to the equations of motion on the optical geometry $(M,g_{ij})$
\begin{align}  
P_{ij} &\equiv R_{ij} - \frac{1}{2}D_i\phi D_j\phi+\frac{1}{3}g_{ij}\frac{\ell^2}{Z^2}V(\phi)+\frac{2}{Z^2}g_{ij}  \nonumber \\
	&= -\frac{2}{Z}D_iD_jZ+\frac{1}{3}g_{ij}\frac{\ell^2}{Z^2}\left(V(\phi)+\frac{6}{\ell^2} \right) \label{Pij}
\end{align}
so that its trace is 
\begin{equation}
P= g^{ij}P_{ij} = R-\frac{1}{2}(D\phi)^2+\frac{\ell^2}{Z^2}\left(V(\phi)+\frac{6}{\ell^2} \right) = \frac{6}{Z^2}(1-(DZ)^2 )  \label{Peqn}
\end{equation}
with the properties
\begin{align}
D_iP &= 6\frac{D^jZ}{Z}\left(\tilde{P}_{ij} - \frac{1}{3}g_{ij}\frac{\ell^2}{Z^2} \left(V(\phi)+\frac{6}{\ell^2} \right)   \right) \label{DP} \\
D^{i}\tilde{P}_{ij} &= \frac{D^iZ}{Z}\tilde{P}_{ij}-\left( \frac{2}{3}\frac{\ell^2}{Z^2}V'(\phi)+D_i\phi\frac{D^iZ}{Z} \right)D_j\phi \label{DPij}
\end{align}
where $\tilde{P}_{ij}$ is the traceless part of $P_{ij}$,
\begin{equation}
\tilde{P}_{ij}  = P_{ij}-\frac{1}{3}g_{ij}P \quad;\quad g^{ij}\tilde{P}_{ij} = 0\quad.
\end{equation}
Taking one more derivative on (\ref{DP}) and using (\ref{phieqn}),(\ref{DPij}) one finds that for static solutions of Einstein-Scalar theory in AlAdS$_4$, the key relation is
\begin{equation}
\boxed{D^2P = -3\tilde{P}_{ij}\tilde{P}^{ij} -\frac{3}{2}\left(D^2\phi\right)^2 - \frac{6\ell^4}{Z^4}\left(\frac{1}{\ell^2}V(\phi)+ \frac{1}{6}V(\phi)^2-\frac{1}{4}V'(\phi)^2 \right)} \label{D2P}
\end{equation}
Since $g_{ij}$ is a Riemannian metric, the first and second term on the right hand side are negative definite while the last term will determine whether $D^2P\leq 0$. The volume integral of $D^2P$ over the whole optical geometry $M$ can be turned into a surface integral at the boundary $\partial M$ by using the divergence theorem. Recall that the boundary $\partial M$ consists of the conformal boundary at infinity $\partial M_{\infty} \simeq \Sigma$, and the killing horizon $\mathcal{H}$ such that $\partial M \simeq \Sigma \cup \mathcal{H}$.
\begin{align}
\int_{M}d\text{Vol}\phantom{.}D^2P &= \int_{\partial M}\star dP \nonumber \\
		&= \int_{\Sigma}\star dP + \int_{\mathcal{H}}\star dP     \label{integrateD2P}
\end{align}
where $\star$ is the Hodge star operator with respect to the optical geometry $(M,g_{ij})$. From the key relation as shown above, $D^2P$ is negative definite only when the last term is negative. Therefore we define the "boundedness function" $B(\phi)$ as the following
\begin{equation}
\boxed{B(\phi) \equiv \frac{1}{\ell^2}V(\phi)+ \frac{1}{6}V(\phi)^2-\frac{1}{4}V'(\phi)^2   }      \label{Bphi}
\end{equation}
so that potentials that satisfy $B(\phi)\geq 0$ (and $B(\phi)=0$ when $\ell^2V(\phi)=-6$) will satisfy the bound $D^2P\leq 0$.
Supposing that the potential satisfies $B(\phi)\geq 0$ then it is guaranteed that $D^2P\leq 0$ everywhere in optical geometry, therefore resulting in the bound on the surface integral over $\partial M$
\begin{equation}
\boxed{\int_{\partial M}\star dP \leq 0 }  \label{intD2P}
\end{equation}
An important question is whether this quantity is finite or not, but before discussing the finiteness and physical applications of this bound let us now discuss potentials that satisfies this bound first.


\subsection{Example of potentials obeying the bound} \label{example}
Since the bound derived above doesn't \textit{a priori} give any clue that says which potentials will satisfy the bound, we have to work it out in the case-by-case basis to verify it.  However, for some generic potentials, we can show that they satisfy the bound in some ranges of their parameters.

\begin{itemize}
\item $\ell^2V(\phi)=-6+\frac{1}{2}m^2\ell^2\phi^2$. The simplest kind of potential with boundedness function:
\begin{equation}
B(\phi) = -\frac{m^2}{4\ell^2}(2+m^2\ell^2)\phi^2+\frac{1}{24}m^4\phi^4
\end{equation}
Therefore $B(\phi) \geq 0$ for relevant and marginal deformations according to $-2 \leq m^2\ell^2 \leq 0$.
\item $\ell^2V(\phi)= -6\cosh{\left(\g\phi\right)}\phantom{a} ; \phantom{a} \g =(-m^2\ell^2/6)^{1/2}$. The generic type of potential that can be obtained from top-down supergravity construction, see \cite{Papadimitriou:2007sj}\cite{Martinez:2004nb} as examples.
\begin{align}
B(\phi) &= \frac{3}{\ell^4}\bigg{(} m^2\ell^2+(4+m^2\ell^2)\cosh{(\g\phi)} \bigg{)}\sinh^2{\left(\frac{\g\phi}{2} \right)} \\
		&= -\frac{m^2}{4\ell^2}(2+m^2\ell^2)\phi^2+\mathcal{O}(\phi^4) \nonumber
\end{align}
Again, $B(\phi) \geq 0$ in the range $-2\leq m^2\ell^2 \leq 0$. This is because the small $\phi$ expansion of any potentials is $\ell^2V(\phi)=-6+\frac{1}{2}m^2\ell^2\phi^2+\mathcal{O}(\phi^3)$ then the small $\phi$ expansion of $B(\phi)$ will always retains the same leading term.

\item $\ell^2V(\phi) = -6+\frac{1}{2}m^2\ell^2\phi^2+\lambda_4\phi^4$, $B(\phi)$ function is
\begin{equation}
B(\phi) = -\frac{m^2}{4\ell^2}(2+m^2\ell^2)\phi^2+\left( \frac{m^4}{24}-\frac{\lambda_4}{\ell^2}-2m^2\lambda_4 \right)\phi^4+\lambda_4\left( \frac{m^2}{6}-4\lambda_4 \right)\phi^6+\frac{1}{6}\lambda_4^2\phi^8
\end{equation}
Since there are two parameters involved, to see which ranges of $m^2, \lambda_4$ make $B(\phi)$ positive for all value of $\phi$ is nontrivial. Nevertheless, we analytically found that $B(\phi) \geq 0$ at the range $\lambda_4 \in [-1/24,0]$ for both $m^2\ell^2=-2$ and 0 cases.
\end{itemize}
Note that even though potentials considered in this section are unbounded from below, they are still "safe" in the sense that their zero temperature solutions are of the "good singularity" type as $\phi \to \infty$ as long as $|V(\phi)|$ grows slower than $\exp(\sqrt{3}\phi)$ as $\phi \to \infty$\cite{Gubser:2000nd}\cite{Kiritsis:2016kog}. Next, we move on to discuss the physical application of the main result (\ref{intD2P}).

\section{Bound on physical quantities} \label{sec4}

Since the main result $\int_{\partial M}\star dP \leq 0$ is already obtained in (\ref{intD2P}), further questions to be answered are: (1) Is it finite? (2) If it is so, what is the physical quantity that this surface integral corresponds to? The answer to the first question can be found by performing an expansion near the surface on which it is integrated. Any suitable near UV or IR coordinates of the bulk spacetime can be translated to optical frame variables and hence $P$ nearby such surfaces can be expressed in terms the optical frame variables as defined in (\ref{Peqn}) and hence we can make sense of the surface integral of $P$. 

\subsection{Finite temperature}  \label{finiteT}
For finite temperature solutions, their bulk geometries end on the killing horizon $\mathcal{H}$ whose surface gravity is related to its Hawking temperature by the relation $\kappa = 2\pi T$ where the surface gravity is defined with respect to the killing vector $\zeta=\partial_{t}$ by $(\zeta\cdot\nabla)\zeta^{A}|_{\mathcal{H}}=\kappa\zeta^{A}|_{\mathcal{H}}$. Here let's consider the near boundary and near horizon expansions of the integrand of the surface integral $\int_{\partial M}\star dP$ and discuss its finiteness.
\begin{itemize}
\item \textbf{Near conformal boundary} \\
Near the conformal boundary $\partial\mathcal{M}$, the metric can be written in the Fefferman-Graham gauge\cite{fefferman2007ambient}\cite{deHaro:2000vlm}:
\begin{equation}
ds^2_{\text{FG}}= \frac{\ell^2}{r^2}dr^2+\g_{\mu\nu}(r,x)dx^{\mu}dx^{\nu}
\end{equation}
the metric itself has the near boundary expansion
\begin{equation}
\g_{\mu\nu}(r,x) = \frac{r^2}{\ell^2}\left( \g_{(0)\mu\nu}+\frac{\ell^4}{r^2}\g_{(2)\mu\nu}+\frac{\ell^{4\Delta_{-}}}{r^{2\Delta_{-}}}\g_{(2\Delta_{-})\mu\nu}+\frac{\ell^6}{r^3}\g_{(3)\mu\nu}+\ldots   \right) 
\end{equation}
where $\g_{(0)\mu\nu}=\bar{g}_{\mu\nu}$ is the metric on conformal boundary and $\g_{(2\Delta_{-})\mu\nu}$ coefficient is a consequence of the backreaction of the scalar field\footnote{To be sufficient for further discussions in this paper for the $\Delta_{-}$ in the range $1 \leq \Delta_{-} < 3/2$ we only need one term that is subleading with respect to $\g_{(2)\mu\nu}$. For $\Delta_{-}=0$ and 1, the contribution from the scalar field is already included in the $\g_{(2)\mu\nu}$ term.}
\begin{equation}
\g_{(0)\mu\nu}dx^{\mu}dx^{\nu} = -dt^2+\bar{g}_{ab}dx^adx^b
\end{equation}
while near boundary expansion of the scalar field takes the same form as (\ref{phiexpansion})
\begin{equation}
\phi(r,x) = \alpha(x)\frac{\ell^{2\Delta_{-}}}{r^{\Delta_{-}}}+\ldots+\beta(x)\frac{\ell^{2\Delta_{+}}}{r^{\Delta_{+}}}+\ldots  
\end{equation}

In the optical frame, $P$ has a near boundary expansion as shown in (\ref{nearboundary})
\begin{equation}
P  = 3\bar{P}(\bar{R},\alpha)-\frac{3}{2}\Delta_{-}\alpha^2\left(\frac{\ell^{2}}{r}\right)^{2\Delta_{-}-2}-6\frac{\mathcal{E}(x)}{C_T}\frac{\ell^2}{r}+\mathcal{O}\left(r^{-2}\right)
\end{equation}
where $\bar{P}$ depends on $\bar{R}$ which is the scalar curvature of the boundary metric $\bar{g}_{\mu\nu}$ and the boundary data of the scalar field, $\alpha$ and $\mathcal{E}(x)=3C_T\g_{(3)tt}$. Hence in order to obtain the surface integral over the boundary, we need the normal derivative of $P$
\begin{equation}
\partial_{n}P = n^r\partial_{r}P = -3\Delta_{-}(\Delta_{-}-1)\alpha^2\left(\frac{\ell^{2}}{r}\right)^{2\Delta_{-}-3}+6\frac{\mathcal{E}(x)}{C_T}+ \mathcal{O}(r^{-1}) \label{dnP}
\end{equation}
Note that the above expression of $\partial_{n}P$ is divergent as $r\to\infty$ thanks to the term $\alpha^2r^{3-2\Delta_{-}}$ for $1 < \Delta_{-} < 3/2$. On the other hand, it is finite in the case of $\Delta_{-}=1,0$ which corresponds to the scalar field mass $m^2\ell^2=-2$ and 0 respectively.
\item \textbf{Near horizon} \\
The near horizon metric in normal radial coordinate whose horizon locates at $r=0$ takes the form\footnote{Note here that the radial coordinate $r$ in near horizon coordinate and $r$ as used in the above FG gauge near $r=\infty$ are generally different coordinates.}
\begin{equation}
\begin{aligned}
ds^2 &= -\kappa^2r^2Q(r,x)dt^2+dr^2+\textit{\textg}_{ab}^{(\mathcal{H})}(r,x)dx^{a}dx^{b}  \\
Q(r,x) &= 1+\mathcal{O}(r^2)  \\
\textit{\textg}_{ab}^{(\mathcal{H})}(r,x) &= \textit{\textg}_{ab}(x) +\mathcal{O}(r^2) 
\end{aligned}
\end{equation}
and the scalar field 
\begin{equation}
\phi(r,x) = \varphi(x)+\mathcal{O}(r^2) 
\end{equation}
where $\kappa =2\pi T$, $\mathcal{R}$ is the scalar curvature of $\textit{\textg}_{ab}$, $\varphi$ is the value of $\phi$ on the horizon and $\mathcal{D}$ is covariant derivative with respect to $\textit{\textg}_{ab}$. 

In the optical frame, with respect to the above near horizon coordinate, a near horizon expansion of $P$ is
\begin{equation}
P=-6\kappa^2+6\kappa^2\left( \frac{1}{\ell^2}+\frac{1}{2}\left(\mathcal{R}-\frac{1}{2}(\mathcal{D}\varphi)^2\right)\right)r^2+\mathcal{O}(r^4)
\end{equation}
\end{itemize}
Using both expansions, given that the divergent term of $\partial_nP$ vanishes, the surface integral (\ref{intD2P}) can be integrated in the following way
\begin{align}
\int_{\partial M}\star dP &= \int_{\Sigma}dA^i \partial_i P+\int_{\mathcal{H}}dA^i \partial_i P  \nonumber \\
			&= \frac{6}{C_T}\int_{\Sigma} \mathcal{E}(x)-24\pi T\left( \frac{A_{\mathcal{H}}}{\ell^2}+2\pi\chi(\mathcal{H})-\frac{1}{4}\int_{\mathcal{H}}(\mathcal{D}\varphi)^2 \right)   \nonumber \\
			&= \frac{6}{C_T}\left( \int_{\Sigma}\mathcal{E}(x)-TS \right)-24\pi T\left(2\pi \chi(\mathcal{H})- \frac{1}{4}\int_{\mathcal{H}}(\mathcal{D}\varphi)^2  \right) \leq 0  \label{boundF}
\end{align}
where entropy is $S=A_{\mathcal{H}}/4G_4=4\pi C_TA_{\mathcal{H}}/\ell^2$. Next, their free energy which is computed from renormalised euclidean on-shell action such that
\begin{equation}
F = TS^{\text{(ren)}}_{\text{E,on-shell}}
\end{equation}
as the details shown in (\ref{Sonshell}).

\paragraph{$\blacktriangleright m^2\ell^2=-2$} : the free energy for the solutions with different boundary conditions are

\begin{enumerate}
\item Dirichlet BC : $\Delta=2$
\begin{equation}
F_{\text{Dirichlet}}(T,\alpha) = \int_{\Sigma}d^2x\sqrt{\bar{g}}\bigg{(}\mathcal{E}(x) -C_T\alpha(x)\beta(x)\bigg{)} -TS \nonumber
\end{equation}
\item Neumann BC : $\Delta=1$
\begin{equation}
F_{\text{Neumann}}(T,\beta) = \int_{\Sigma}d^2x\sqrt{\bar{g}}\mathcal{E}(x) -TS \nonumber 
\end{equation}
\item General BC(or Robin BC) : $\Delta=1, \beta=\frac{\partial W}{\partial \alpha}$ the free energy can be calculated as outlined in \cite{Anabalon:2015xvl}
\begin{equation}
F_{\text{Robin}}(T,\alpha) =\int_{\Sigma}d^2x\sqrt{\bar{g}}\bigg{(}\mathcal{E}(x) -C_T\alpha(x)\beta(x)+W(\alpha) \bigg{)} -TS \nonumber
\end{equation}
\end{enumerate}

\paragraph{$\blacktriangleright m^2\ell^2=0$} : $\Delta=3$, only Dirichlet BC can be applied 
\begin{equation}
F_{m^2=0}(T) = \int_{\Sigma}d^2x\sqrt{\bar{g}}\mathcal{E}(x)-TS \nonumber
\end{equation}

\paragraph{}As a consequence of the inequality (\ref{boundF}), the free energy can be bounded from above
\paragraph{$\blacktriangleright m^2\ell^2=-2$}
\begin{enumerate}
\item Dirichlet BC 
\begin{equation}
F_{\text{Dirichlet}} \leq 4\pi TC_T\left( 2\pi \chi(\mathcal{H})-\frac{1}{4}\int_{\mathcal{H}}(\mathcal{D}\varphi)^2 \right)- C_T\int_{\Sigma}d^2x\sqrt{\bar{g}}\alpha(x)\beta(x)
\end{equation}
\item Neumann BC 
\begin{equation}
F_{\text{Neumann}} \leq 4\pi TC_T\left( 2\pi \chi(\mathcal{H})-\frac{1}{4}\int_{\mathcal{H}}(\mathcal{D}\varphi)^2 \right)
\end{equation}
\item Robin BC
\begin{equation}
F_{\text{Robin}}  \leq 4\pi TC_T\left( 2\pi \chi(\mathcal{H})-\frac{1}{4}\int_{\mathcal{H}}(\mathcal{D}\varphi)^2 \right)- \int_{\Sigma}d^2x\sqrt{\bar{g}}\bigg{(}C_T\alpha(x)\beta(x)-W(\alpha) \bigg{)} 
\end{equation}
\end{enumerate}

\paragraph{$\blacktriangleright m^2\ell^2=0$}
\begin{equation}
F_{m^2=0} \leq 4\pi TC_T\left( 2\pi \chi(\mathcal{H})-\frac{1}{4}\int_{\mathcal{H}}(\mathcal{D}\varphi)^2 \right)
\end{equation}
The above result for $m^2=0$ can be seen as a generalisation of the bound presented in \cite{Fischetti:2016vfq} which is only valid for the scalar field without nontrivial higher order terms in the potential, while the bound presented here is valid for any potential regardless of higher order terms in $\phi$ as long as the potential satisfy $B(\phi)\geq 0$.
Having obtained these results, we can move on to consider the bound for zero temperature solutions.


\subsection{Zero temperature}  \label{zeroT}
Having obtained the finite temperature result for AlAdS$_4$ with the conformal boundary $\partial \mathcal{M}\simeq \mathbb{R}\times \Sigma$ where $\Sigma$ has finite volume, we are able to consider the zero temperature vacuum state of the bulk geometry (and hence of the holographic CFT$_3$ as well). As we are considering the vacuum solutions that can be taken from zero temperature limit of black hole solutions \cite{Gubser:2000nd}, such spacetimes have zero or finite entropy as $T\to 0$\footnote{An example of a spacetime with $S\to 0 $ as $T\to 0$ is the zero temperature limit of toroidal black hole which ends on the quotient of the singular Poincare horizon in the IR\cite{Cheamsawat:2019gho}. For a spacetime with finite entropy as $T\to 0$, an example is the zero temperature limit of topological black hole\cite{Emparan:1999gf} whose bulk geometry ends on the degenerate horizon.}(such that the $TS$ term in the free energy vanishes at $T=0$, $\lim_{T\to 0}TS = 0$) the vacuum energy bound can be obtained as a zero temperature limit of the free energy bound
\begin{equation}
E_{\text{vac}} = \lim_{T\to 0}F(T)
\end{equation}
Hence for conformal dimensions $\Delta=1,2,3$ as discussed previously, the vacuum energy bound becomes
\begin{equation}
E_{\text{vac}} \leq -C_T\int_{\Sigma}d^2x\sqrt{\bar{g}}\alpha(x)\beta(x) \quad;\quad \text{for } \Delta=2  
\end{equation}
\begin{equation}
E_{\text{vac}} \leq 0 \quad;\quad \text{for } \Delta=1,3   
\end{equation}
\begin{equation}
E_{\text{vac}} \leq -\int_{\Sigma}d^2x\sqrt{\bar{g}}\left(C_T\alpha(x)\beta(x)-W(\alpha) \right) \quad;\quad \text{for } \Delta=1 \text{ with Robin BC}
\end{equation}

\paragraph{Generalisation to infinite volume boundary} Even though our main focus is the case where the spatial part of conformal boundary $\Sigma$ has finite volume, under some restrictions the bound on free energy and vacuum energy can be generalised to the case of infinite volume boundary in the following senses.
\begin{itemize}
\item Holographic lattices : spatially periodic source fields on the conformal boundary\footnote{Such a deformation would correspond to the sources that take the form $\alpha(x)\sim \cos{kx}, \bar{g}_{ab}\sim \cos{kx}$.} have great applicability in the context of holographic lattices(such as \cite{Horowitz:2012ky}\cite{Donos:2013eha}). They are normally realised on noncompact spaces. Our result for $\Sigma \simeq T^2$ can be decompactified and then the bounds as described above can be seen as the bound of such quantities \emph{per unit cell} of the lattice.

\item CFT$_3$ with a localised deformation : for the deformations where the sources $\alpha, \bar{g}_{ab}$ are localised (see e.g.\cite{Horowitz:2014gva}\cite{Janik:2015oja} for numerical solutions and the analysis in this setting). In the zero temperature case, we \textit{speculate} that if the sources on the conformal boundary decay rapidly enough (for example, in the case of chemical potential source in Einstein-Maxwell theory as in \cite{Horowitz:2014gva}), the IR geometry would end on the null Poincare horizon\cite{Hickling:2014dra}. This horizon is regular and the near horizon expansion is smooth there (an example for pure gravity case is shown in appendix.(\ref{nearextremalhorizon})) If this scenario is true, $\int_{\partial M}\star dP$ will not receive any contribution from the IR geometry and the region far away from the localised source on the conformal boundary. Hence the vacuum energy bound can be seen as the Casimir energy bound due to the localised deformations.
\end{itemize}


\section{Reverse engineering the potential}   \label{sec5}
\paragraph{}The logic that we followed until now is that for Einstein-scalar theory with arbitrary potential $V(\phi)$ with mass parameter $m^2\ell^2=-2,0$ whether there exists a bound on energy or free energy can be tested by putting $V(\phi)$ into equation (\ref{Bphi}) and checking if $B(\phi)\geq 0$ or not. If $B(\phi)\geq 0$, it is ensured that there exists the upper bound of the form as shown in section (\ref{finiteT}) and (\ref{zeroT}).
\paragraph{}The logic in this section is different, equation (\ref{Bphi}) that is defining $B(\phi)$ can be viewed in another way as a differential equation to be solved for $V(\phi)$ with a source given by an arbitrary function $B(\phi)$ 
\begin{equation}
V'(\phi)^2 = 4\left(\frac{1}{\ell^2}V(\phi)+\frac{1}{6}V(\phi)^2-B(\phi) \right)     \label{ODE}
\end{equation}
For a given positive $B(\phi)$ function, $V(\phi)$ can be solved from this equation with initial condition $V(0)=-6/\ell^2$. Then it can be plugged into the Einstein equation and scalar field equation (\ref{main}) to solve for $g_{AB}$ and $\phi$ which results in the solution whose vacuum energy or free energy satisfies the upper bound.
\paragraph{}In the simplest case when $B(\phi)$ is identically zero, the differential equation (\ref{ODE}) can be directly integrated to obtain the nontrivial solution which is not $V(\phi)=-6/\ell^2$. The solution is
\begin{equation}
\ell^2V_{B=0}(\phi) = -3-3\cosh{\left(\sqrt{2/3}\phi \right)} = -6-\phi^2+\ldots
\end{equation}
In the next section, we will use this potential to demonstrate how the bound is satisfied for both finite and zero temperature solutions.


\section{Consistency check from spatially homogeneous solutions}   \label{sec6}
In this section, we will examine a detailed example for the scalar field with mass $m^2\ell^2=-2$. Using the  example of $\ell^2V(\phi)=-3-3\cosh{\left(\sqrt{2/3}\phi \right)}$. In the spherically symmetric setting we consider black hole and soliton solution where the scalar field is subjected to a generalised boundary condition,
\begin{equation} 
\phi(r) = \alpha \frac{\ell^2}{r}+f\alpha^2\frac{\ell^4}{r^2}+\ldots \label{scalardesigner}
\end{equation}  
where the coefficient $\beta=f\alpha^2$ is characterised by the function $W(\alpha)$ such that
\begin{equation}
\beta = f\alpha^2 =\frac{ \partial W(\alpha)}{\partial \alpha}  \quad \Rightarrow \quad W(\alpha) = \frac{1}{3}f \alpha^3
\end{equation}
This boundary condition and $W(\alpha)$ function is holographically dual to the deformation with operator of conformal dimension $\Delta=\Delta_{-}=1$ where $\alpha$ and $W(\alpha)$ are identified with the deformation of the holographic CFT in the following way
\begin{equation} 
\alpha \sim \langle \mathcal{O} \rangle \quad ,\quad S_{\text{CFT}} \to S_{\text{CFT}} + \int d^3x\sqrt{\bar{g}}\frac{f}{3}\mathcal{O}^3
\end{equation}
which is the triple trace deformation preserving conformal symmetry in field theory\footnote{In general dimensions, the multitrace deformation preserving conformal symmetry takes the form $W(\alpha) = k\alpha^{d/\Delta_{-}}$ where $k$ is constant.}.
\paragraph{}In both finite and zero temperature cases, we take the spherically symmetric metric ansatz of the form
\begin{equation}
ds^2 = -g(r)e^{-2\chi(r)}dt^2 + g(r)^{-1}dr^2+r^2d\Omega_2^2 \label{ansatz}
\end{equation}
With this ansatz, Einstein equation and scalar field equation take the form(in $\ell=1$ unit)
\begin{equation} \label{eom}
\begin{aligned}
0 &= rg'(r)+g(r)\left(1+\frac{1}{4}r^2\phi'(r)^2 \right)-1+\frac{1}{2}r^2V(\phi) \\
0 &= \chi'(r)+\frac{1}{4}r\phi'(r)^2 \\
0 &= g(r)\phi''(r)+g(r)\phi'(r)\left( \frac{2}{r}+\frac{r}{4}\phi'(r)^2 \right)+g'(r)\phi'(r)-V'(\phi) 
\end{aligned}
\end{equation}
From now on in this section, we will take $\ell=1$ and $16\pi G_4=1=C_T$ for simplicity.

\subsection{Black hole solution}
For the black hole solution, apart from the scalar field as in (\ref{scalardesigner}), the asymptotic behaviour of metric functions $g(r)$ and $\chi(r)$ take the from
\begin{align}
g(r) &= r^2 +\left(1+\frac{\alpha^2}{4}\right)-\frac{\mu}{r}+\mathcal{O}\left(\frac{1}{r^2} \right) \\
\chi(r) &= \chi_{0}+\frac{\alpha^2}{8r^2}+ \mathcal{O}\left(\frac{1}{r^3} \right)
\end{align}
Furthermore, at the horizon $r=R_{h}$ such that $g(R_{h})=0$, the near horizon expansion for $g(r), \chi(r), \phi(r)$ take the form
\begin{align}
g(r) &= g'(R_{h})(r-R_{h})+\mathcal{O}((r-R_{h})^2)  \\
\chi(r) &= \chi_{h} +\mathcal{O}(r-R_{h}) \\
\phi(r) &= \phi_{h}+\mathcal{O}(r-R_{h}) 
\end{align}
where $g'(R_{h})$ depends on the horizon radius and scalar field at the horizon ($R_{h}$ and $\phi_{h}$)
\begin{equation}
g'(R_{h}) = \frac{1}{R_{h}}-\frac{R_{h}}{2}V(\phi_{h})
\end{equation}
For each fixed $f$ of the scalar field boundary condition, there are $\{ \mu, \alpha, \chi_0 \}$ as the boundary data and $\{ R_{h}, \phi_{h}, \chi_{h} \}$ as the horizon data. However, $\chi_{0}$ can always be set to zero by redefinition of the time coordinate: $t \to te^{\chi_0}$ so that $\chi_{h}$ can be fixed at an arbitrary value.

\begin{figure}[h!]
\centering
\includegraphics[width=0.5\textwidth]{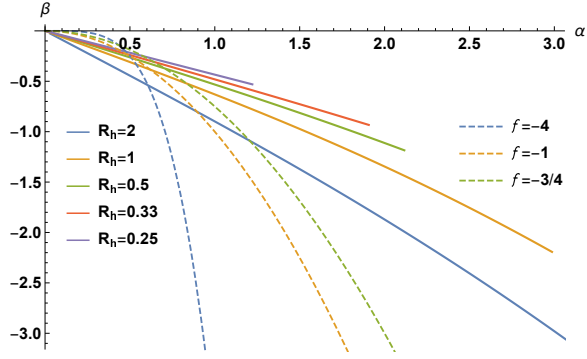}
\caption{$\beta_{R_{h}}(\alpha)$ curves(solid curves) for black hole solutions with different horizon radii $R_{h}$. Along each curve is a one-parameter family of solutions parametrised by $\phi_{h}$, the allowed black hole solution for each boundary condition $\beta_{\text{bc}}(\alpha)$(dashed curves) is identified with the point where $\beta_{\text{bc}}(\alpha)$ curve intersects $\beta_{R_{h}}(\alpha)$ curve.}
\label{BHcurve}
\end{figure}

\begin{figure}[h!] 
\centering
\includegraphics[width=0.5\textwidth]{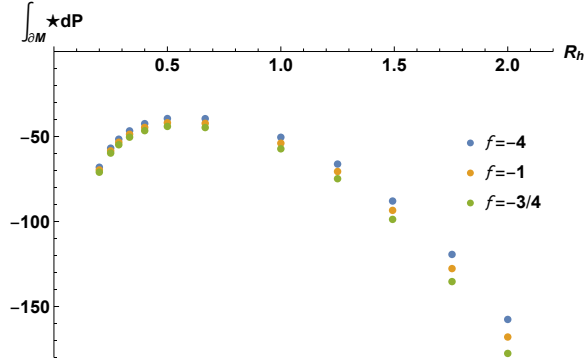}
\caption{Values of $\int_{\partial M}\star dP$ for black hole solutions in designer gravity of various horizon radii for scalar field boundary condition $\beta = f\alpha^2$ for $f\in \{-4,-1,-3/4\}$.}
\label{designerBHbound}
\end{figure}

\paragraph{Thermodynamics quantities}for these given boundary and horizon data, thermodynamics quantities of the black hole can be calculated as the following
\begin{itemize}
\item Mass :  The mass of black hole can be calculated from the holographic renormalisation procedure, and for designer gravity with $m^2=-2$, boundary counterterms are given in \cite{Anabalon:2015xvl}. The mass of the black hole is
\begin{align}
M &= 4\pi (2\mu+\alpha\beta+W(\alpha)) 
\end{align} 
or in terms of $\mathcal{E}$,
\begin{equation}
M = 4\pi (\mathcal{E}-\alpha\beta + W(\alpha))
\end{equation}
\item Temperature
\begin{equation}
T = \frac{1}{4\pi}e^{-\chi_{h}} g'(R_{h})
\end{equation}
\item Entropy
\begin{equation}
S = \frac{A_{\mathcal{H}}}{4G_4} = 16\pi^2R_{h}^2
\end{equation}
\end{itemize}
To verify the holographic bound (\ref{intD2P}) at finite termperature, we look at the quantity (\ref{boundF}). In this case, $\mathcal{D}\varphi = 0$ and $\chi(\mathcal{H})=\chi(S^2)=2$, we have to check the inequality 
\begin{equation}
\int_{S^2}\star dP = 4\pi \mathcal{E}-ST-16\pi^2T \leq 0 \label{BHbound}
\end{equation}
we show numerical results in Fig.(\ref{designerBHbound}) for a sample of black hole solutions which are obtained using the method outlined in \cite{Hertog:2006rr}(see Fig.(\ref{BHcurve})).
It is obvious that the bound is satisfied for all solutions considered. Following the holographic renormalisation with counterterm action as presented in \cite{Anabalon:2015xvl}, the free energy (in terms of $\mathcal{E}$) of black hole solutions is
\begin{equation}
F = 4\pi \left(\mathcal{E}-\alpha\beta+W(\alpha) \right)-TS
\end{equation}
Then, as a consequence of the bound (\ref{BHbound}), the free energy upper bound of black hole solutions in designer gravity at $m^2=-2$ is 
\begin{equation}
F \leq 16\pi^2T+4\pi \left( W(\alpha)-\alpha\beta \right)
\end{equation}


\subsection{Soliton solution}
For soliton solutions, asymptotic expansions of $g(r), \chi(r), \phi(r)$ are the same. Since soliton solutions are regular, then the near origin($r=0$) the expansion takes the form
\begin{align} 
g(r) &= 1-\frac{1}{6}V(\phi_c)r^2+\mathcal{O}(r^4) \\
\chi(r) &= \chi_{c}-\frac{1}{144}V'(\phi_c)^2r^4+\mathcal{O}(r^6) \\
\phi(r) &= \phi_{c}+\frac{1}{6}V'(\phi_{c})r^2+\mathcal{O}(r^4) 
\end{align}
and since $\chi_{c}$ can be fixed to be an arbitrary value, then soliton solutions are a one-parameter family of solutions parametrised by $\phi_{c}$. Therefore, at each value of $\phi_{c}$, the equations of motion (\ref{eom}) can be integrated from $r=0$ to arbitrarily large $r$ and boundary data $\mu, \alpha, \beta$ can be obtained for each solution \cite{Hertog:2004ns}.
As for black hole solutions, soliton solutions have $\mathcal{E}=2(\mu+\alpha\beta)$. Therefore, as a consistency check of the holographic bound (\ref{intD2P}), we have to check the bound
\begin{equation}
\int_{S^2} d\Omega\phantom{.}\mathcal{\mathcal{E}} = 4\pi \mathcal{E} \leq 0
\end{equation}
\begin{figure}[h!] 
\begin{center}
\begin{tabular}{cc}
\begin{subfigure}{0.47\textwidth}
\includegraphics[width=\textwidth]{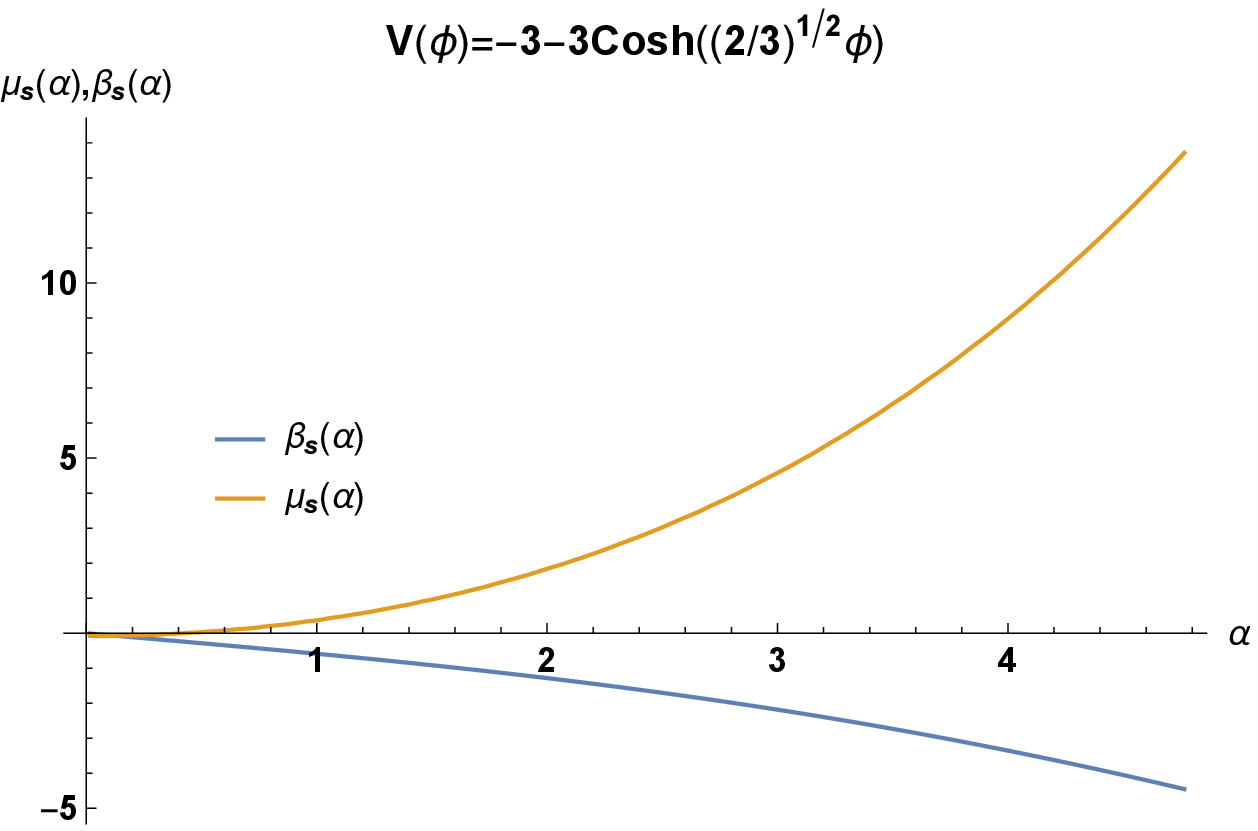} 
\end{subfigure}
  & \begin{subfigure}{0.47\textwidth}
\includegraphics[width=\textwidth]{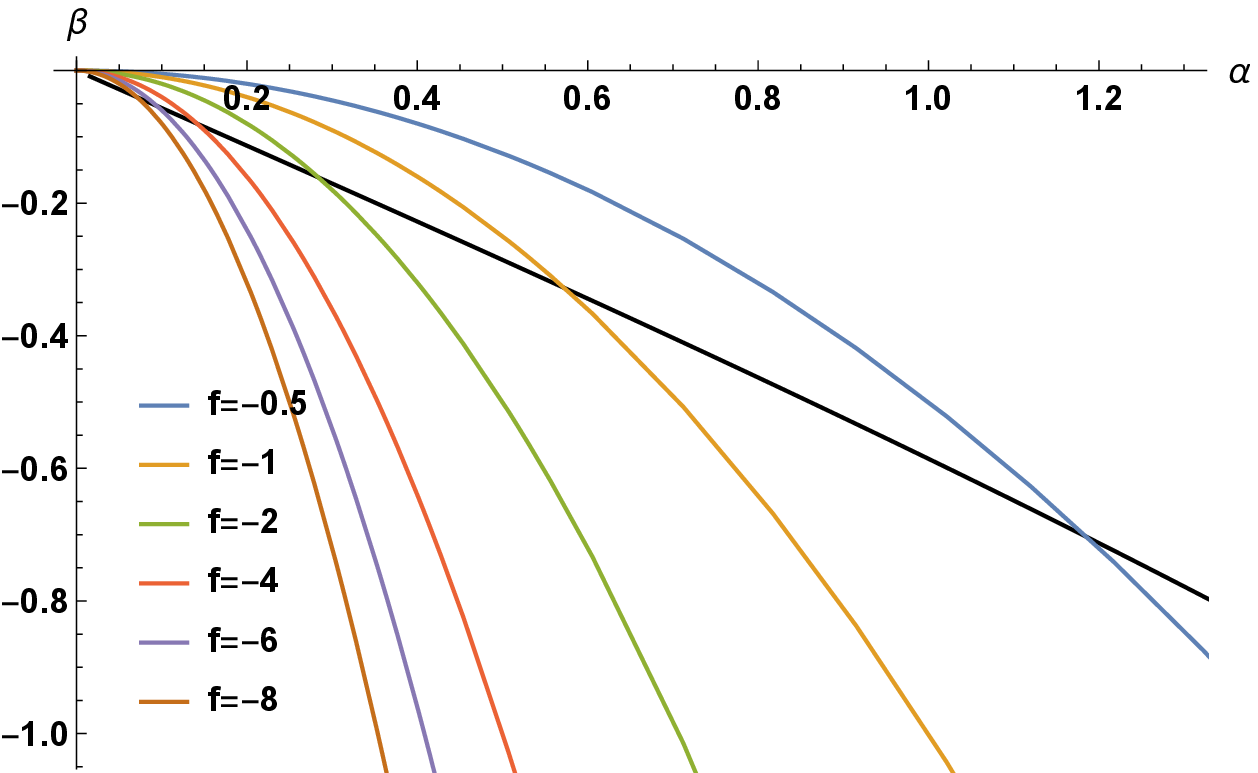} 
\end{subfigure}
\end{tabular}
\end{center}
\caption{\textbf{Left}: $\mu$ and $\beta$ for soliton solutions. \textbf{Right}: $\beta_s(\alpha)$ for soliton solution(black curve) where each ($\alpha, \beta_{s}$) point along the curve correspond to one value of $\phi_{c}$. The allowed soliton solution for each boundary condition $\beta_{\text{bc}}=f\alpha^2$(colored curves) is identified with the point where $\beta_{\text{bc}}(\alpha)=\beta_{s}(\alpha)$.}
\label{solitoncurve}
\label{solitonpic}
\end{figure}
\begin{figure}[h!] 
\centering
\includegraphics[width=0.5\textwidth]{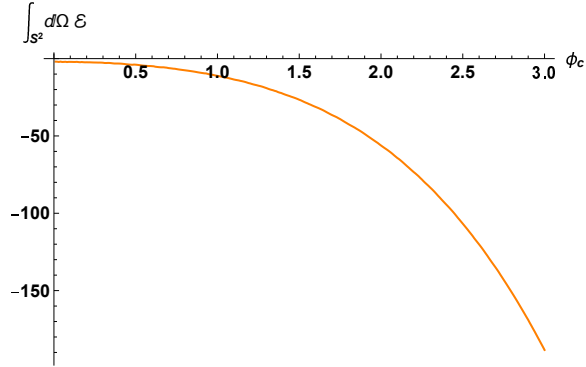}
\caption{$4\pi \mathcal{E}$ of soliton solutions specified by $\phi_c$.}
\label{solitonbound}
\end{figure}
From Fig.(\ref{solitonbound}), soliton solutions for the scalar field with $V(\phi) = -3-3\cosh{(\sqrt{2/3}\phi )})$ obviously satisfy the bound (\ref{intD2P}). Then, as a consequence of this bound, the soliton mass 
\begin{equation}
M_s = 4\pi (\mathcal{E}-\alpha\beta+W(\alpha))
\end{equation}
is bounded from above. This result is an upper bound as opposed to the positive energy bound for spherically symmetric solitons in designer gravity which is the lower bound as has been advocated in \cite{Faulkner:2010fh}\cite{Amsel:2007im}\cite{Hertog:2005hm}\cite{Amsel:2006uf}. We emphasise that this lower bound can be applied to more general scalar potentials but only for spherically symmetric soliton solutions and hence the bulk is asymptotically global AdS$_{4}$ with a constant scalar source. Therefore, in the cases where $B(\phi)\geq 0$ is satisfied, the mass of spherically symmetric solitons is bounded from both above and below
\begin{equation}
4\pi W(\alpha) \leq M_s \leq 4\pi (W(\alpha)-\alpha\beta)
\end{equation}
Note that this also implies that $\alpha\beta$ is indeed negative for such spherical symmetric solutions as the numerical result shown in Fig.(\ref{solitonpic}) confirms.


\section{Conclusion \& Discussion}   \label{sec7}
\paragraph{Summary}In this work, we derived the upper bound on physical quantities, namely vacuum energy and free energy for static solutions in Einstein-Scalar theory in four dimensional AlAdS spacetime. A simple criterion to test whether the Einstein-Scalar system with arbitrary potentials $V(\phi)$ with the mass parameter $m^2\ell^2=-2,0$ obey the bound is given in terms of the function $B(\phi)$ which is determined purely by the potential of the scalar field. Some examples of potentials that can satisfy the bound are shown and studied.
\paragraph{}The upper bound on vacuum energy for solutions at zero temperature as shown in (\ref{zeroT}) is only dependent on the boundary data of the scalar field while the upper bound on free energy of finite temperature solutions as shown in (\ref{finiteT}) is determined by the boundary data as well as the horizon data of the scalar field and the topology of the killing horizon where the spacetime in the bulk ends(assuming the geometry is regular there). 

\paragraph{Implications for QFTs}The results shown in this paper so far are stated in terms of the upper bound on energetic quantities for the bulk theory in an AlAdS spacetime. However, as a consequence of holographic duality, the energy and free energy bound can be mapped directly to the vacuum energy and free energy of holographic CFT in static curved spaces, $d\bar{s}^2=-dt^2+d\Sigma^2$ which is deformed by the dual operator $\mathcal{O}$ as $S_{\text{CFT}}\to S_{\text{CFT}}+\int d^3x\sqrt{\bar{g}}J(x)\mathcal{O}(x)$ where $J(x)$ is the inhomogeneous function on the curved space\footnote{$J(x)$ becomes coupling constant in the case that it is constant value.}. One related question is this respect is the field theoretical meaning of the boundedness function $B(\phi)$ which seems to be obscure in the sense that it is not known whether there are something special about it in terms of field theory. Is it just something that happens to be "useful" from the analysis of equations of motion?

\paragraph{Open questions}Apart from those we already discussed, there are some open questions arising from our results;

\begin{itemize}

\item Bounding $\int_{\Sigma}\alpha\beta$ : The bound derived earlier is the bound on the integral of $\mathcal{E}$ which basically is the Fefferman-Graham coefficient of $g_{tt}$ which is dual to (part of) $\langle T_{tt} \rangle$ of the holographic CFT. Is there any way to bound $\int_{\Sigma}\alpha\beta$ by more fundamental quantities of the spacetime?

\item Generalisation to other $\Delta$ : The approach shown in this paper can only obtain the result for the dual operator of dimensions $\Delta=1,2,3$ since the divergent term from the near conformal boundary part of the surface integral $\int_{\partial M}\star dP$ (\ref{dnP}) vanishes only at these values of $\Delta$. Is there a more general approach or divergence cancellation which yields the bound for relevant deformations of general $\Delta$?

\item Inclusion of other fields in the bulk : An obvious direction that can be extended from this work is finding the bound on physical quantities of other theories in the bulk where extra fields apart from the pure gravity are included such as multiple scalar fields, Maxwell field, extra $U(1)$ gauge fields, etc. If such a bound could be found in these types of theories, it could be valuable in the context of AdS/CMT correspondence in which many holographic models for condensed matter systems are based on Einsteins-Maxwell theory such as Einstein-Maxwell-Dilaton model\cite{Charmousis:2010zz}\cite{Goldstein:2009cv}\cite{Cadoni:2009xm}, Einteins-Maxwell-Abelian Higgs model\cite{Hartnoll:2008kx}\cite{Gauntlett:2009dn}\cite{Bhaseen:2012gg} and the extensions thereof.

\item Relation to the Gauss-Bonnet theorem : For pure gravity in AlAdS$_4$, the free energy bound in \cite{Hickling:2015tza} can also be derived from a purely geometric way(without referring to physical quantities in AlAdS) in terms of the "renormalised volume bound" which is proven by the use of the Gauss-Bonnet theorem \cite{anderson20002} without restricting to only  static solutions. Then the obvious question is: can the result presented in this paper be derived from the generalisation of the 'Gauss-Bonnet theorem' method of proof?

\end{itemize}


\acknowledgments
The author(KC) would like to thank Toby Wiseman for suggesting this problem, a lot of useful discussions, carefully reading the manuscript and encouragement throughout the duration of this project. KC would also like to thank Matthew Roberts for insightful explanation and bringing the literatures on designer gravity to attention, Igal Arav and Jerome Gauntlett for comments and useful discussions. KC has been sponsored by DPST scholarship from the Royal Thai Government.


\appendix  
\section{Near boundary and near horizon expansion in optical frame} \label{app}
WARNING: In this section, the notation "$r$" is used repetitively to represent the radial coordinate in different situations we consider in each subsections. To prevent further confusions, we note here that $r \to \infty$ as we reach the conformal boundary of AlAdS and $r \to 0$ as we reach any horizon(extremal or non-extremal) deep inside the bulk. Moreover, the coordinate "$r$" we used in each subsection are not necessarily defined in the same coordinate patch, nor the extendable to the whole spacetime with a single coordinate patch.

\subsection{Near boundary expansion} \label{nearboundary}
For the metric in optical frame as shown in (\ref{optical}), we can define the radial coordinate $r$ such that the conformal boundary is located at $r=\infty$. Near the conformal boundary, a metric of AlAdS spacetime can be written in the Fefferman-Graham gauge in which we can perform the near boundary expansion of the metric and scalar field efficiently. We arrange the metric in Fefferman-Graham gauge such that
\begin{equation}
ds^2 = \frac{\ell^2}{r^2}dr^2+\g_{\mu\nu}(r,x)dx^\mu dx^{\nu}
\end{equation}
In this detailed calculation, we will show the result for cases where the mass of the scalar field lies in the range $m_{BF}^2\ell^2 < m^2\ell^2 < m_{BF}^2\ell^2+1$ and $m^2\ell^2=0$ respectively.
In FG gauge $\g_{\mu\nu}$ and the scalar field have the asymptotic expansion in both cases as follows:
\begin{itemize}
\item $m_{BF}^2\ell^2 < m^2\ell^2 < m_{BF}^2\ell^2+1$ : In particular, the case of $m^2\ell^2=-2$ that we consider in the main text lies in the smaller subrange $m_{BF}^2\ell^2 < m^2\ell^2 \leq m_{BF}^2\ell^2+1/4$ such that $1 \leq 
\Delta_{-} < 3/2$ , $3/2  < \Delta_{+} \leq 2$. In the absence of cubic interactions in $V(\phi)$\footnote{If the cubic interaction term is present, the near boundary expansion of $\phi$ with mass $m^2\ell^2 =-2$ will have the logarithmic term\cite{Henneaux:2006hk}. For simplicity, we will not consider such a case here.}
\begin{align}
\g_{\mu\nu}(r,x) &= \frac{r^2}{\ell^2}\left( \g_{(0)\mu\nu}+\frac{\ell^2}{r^2}\g_{(2)\mu\nu}+\frac{\ell^{4\Delta_{-}}}{r^{2\Delta_{-}}}\g_{(2\Delta_{-})\mu\nu}+\frac{\ell^6}{r^3}\g_{(3)\mu\nu}+\ldots   \right) \\
\phi(r,x) &= \alpha(x)\frac{\ell^{2\Delta_{-}}}{r^{\Delta_{-}}}+\beta(x)\frac{\ell^{2\Delta_{+}}}{r^{\Delta_{+}}}+\ldots
\end{align}
where the term $\g_{(2\Delta_{-})\mu\nu}$ is the consequence of the backreaction of the scalar field which is of order $\mathcal{O}(\phi^2)$\footnote{For other values of mass outside of the range $m_{BF}^2\ell^2 < m^2\ell^2 \leq m_{BF}^2\ell^2+1/4$ there could be other subleading terms such as $r^{-3\Delta_{-}},r^{-4\Delta_{-}}$ which are of order $\alpha^3, \alpha^4$ respectively. For the detailed analysis, see \cite{Henneaux:2006hk}.}. Coefficients $\g_{(2)\mu\nu}, \g_{(2\Delta_{-})\mu\nu}$ can be expressed in terms of boundary data of the metric and scalar field as
\begin{equation}
\begin{aligned}
\g_{(2)\mu\nu} &= -\left(R_{\mu\nu}(\g_{(0)})-\frac{1}{4}\g_{(0)\mu\nu}R(\g_{(0)}) \right)-\frac{1}{8}\alpha^2\g_{(0)\mu\nu}\quad; \quad \Delta_{-}=1   \\
\g_{(2)\mu\nu} &= -\left(R_{\mu\nu}(\g_{(0)})-\frac{1}{4}\g_{(0)\mu\nu}R(\g_{(0)}) \right) \quad; \quad  1< \Delta_{-} < 3/2    \\
\g_{(2\Delta_{-})\mu\nu}  &= -\frac{1}{8}\alpha^2\g_{(0)\mu\nu} \quad; \quad  1< \Delta_{-} < 3/2   
\end{aligned}
\end{equation}
and the trace of $\g_{(3)\mu\nu}$ is
\begin{align}
\g_{(3)} &= \g^{\mu\nu}_{(0)}\g_{(3)\mu\nu} = -\frac{2}{3}\Delta_{-}(\Delta_{+}-\Delta_{-})\alpha\beta  \quad; \quad  1\leq \Delta_{-} < 3/2 \\
\g_{(3)} &= 0  \quad; \quad  \Delta_{-} =0 \quad (m^2=0)
\end{align}
\item $m^2\ell^2=0$
\begin{align}
\g_{\mu\nu}(r,x) &= \frac{r^2}{\ell^2}\left( \g_{(0)\mu\nu}+\frac{\ell^2}{r^2}\g_{(2)\mu\nu}+\frac{\ell^6}{r^3}\g_{(3)\mu\nu}+\ldots   \right) \\
\phi(r,x) &= \alpha(x)+\frac{1}{2}(\square_{(0)}\alpha(x))\frac{\ell^4}{r^2}+\beta(x)\frac{\ell^{6}}{r^3}+\ldots
\end{align}
where $\square_{(0)}$ is the d'Alembertian with respect to $\g_{(0)\mu\nu}$. $\g_{(2)\mu\nu}$ is also taking the backreaction of the scalar field into account
\begin{equation}
\g_{(2)\mu\nu} = -\left(R_{\mu\nu}(\g_{(0)})-\frac{1}{4}\g_{(0)\mu\nu}R(\g_{(0)}) \right)+\frac{1}{2}\left( \partial_{\mu}\alpha\partial_{\nu}\alpha -\frac{1}{4}\g_{(0)\mu\nu}(\partial\alpha)^2 \right)
\end{equation}
furthermore, the trace of $\g_{(3)\mu\nu}$ vanishes, $\g_{(3)}=0$.
\end{itemize}
According to the setting considered in the paper $\g_{(0)\mu\nu}=\bar{g}_{\mu\nu}$ in (\ref{boundary}). These near boundary expansions for the metric and scalar field can be translated to the optical frame variables as the following: 
\begin{equation}
\frac{\ell^2}{Z^2}\left(-dt^2+g_{ab}dx^adx^b+g_{rr}dr^2 \right)=\g_{tt}dt^2+\g_{ab}dx^adx^b+\frac{\ell^2}{r^2}dr^2 \nonumber 
\end{equation}
This leads to FG expansions for $Z, g_{ab}$ and $g_{rr}$(We show $\g_{(2\Delta_{-})}$ in the general expressions. For $m^2\ell^2=0$, this $\g_{(2\Delta_{-})}$ term should be omitted.)
\begin{equation}
\begin{aligned}
Z(r,x^a) &= \frac{\ell^2}{r}\left(1+\g_{(2)tt}\frac{\ell^4}{2r^2}+ \g_{(2\Delta_{-})tt}\frac{\ell^{4\Delta_{-}}}{2r^{2\Delta_{-}}} +\g_{(3)tt}\frac{\ell^6}{2r^3}+\ldots \right)   \\
g_{rr}(r,x^a) &= \frac{\ell^4}{r^4}\left(1+\g_{(2)tt}\frac{\ell^4}{r^2}+ \g_{(2\Delta_{-})tt}\frac{\ell^{4\Delta_{-}}}{r^{2\Delta_{-}}} + \g_{(3)tt}\frac{\ell^{6}}{r^3} +\ldots \right) \\
g_{ab}(r,x^a) &= \g_{(0)ab}+(\g_{(2)ab}+\g_{(0)ab}\g_{(2)tt})\frac{\ell^4}{r^2}+ (\g_{(2\Delta_{-})ab}+\g_{(0)ab}\g_{(2\Delta_{-})tt})\frac{\ell^{4\Delta_{-}}}{r^{2\Delta_{-}}}  \\
		&\phantom{=} +(\g_{(3)ab}+\g_{(0)ab}\g_{(3)tt})\frac{\ell^6}{r^3}+\ldots \\
g_{ra}(r,x^a) &= 0
\end{aligned}
\end{equation} 
where $\g_{(0)tt}=-1$ and $\g_{(0)ab}=\bar{g}_{ab}$.
Then the near boundary expansion of $P$ is
\begin{align}
P &= -12\g_{(2)tt}-12\Delta_{-}\g_{(2\Delta_{-})tt}\left(\frac{\ell^{2}}{r}\right)^{2\Delta_{-}-2}-18\g_{(3)tt}\frac{\ell^2}{r}+\ldots  \nonumber \\
	&= -12\g_{(2)tt}-\frac{3}{2}\Delta_{-}\alpha^2\left(\frac{\ell^{2}}{r}\right)^{2\Delta_{-}-2}-6\frac{\mathcal{E}(x)}{C_T}\frac{\ell^2}{r}+\ldots
\end{align}
where we define $\mathcal{E}(x) = 3C_T\g_{(3)tt}$ as a part from the metric that always contributes to energy density in the holographic stress tensor $\langle T_{\mu\nu} \rangle$. Then the conformal boundary contribution to the surface integral of $\partial_{i}P$ (in optical frame) becomes
\begin{align}
\int_{\partial M_{\infty}}dA^{i}\partial_{i}P &= \int_{\partial M_{\infty}}d^2x \sqrt{g_{(2)}}n^r\partial_r P \nonumber 
\end{align}
where 
\begin{align}
n^r &= \frac{1}{\sqrt{g_{rr}}} = \frac{r^2}{\ell^2}(1+\mathcal{O}(r^{-2})) \nonumber \\
\sqrt{g_{(2)}} &= \sqrt{\bar{g}}\left(1+\mathcal{O}(r^{-2}) \right)    \nonumber \\
n^r\partial_{r}P &= -3\Delta_{-}(\Delta_{-}-1)\alpha^2\left(\frac{\ell^{2}}{r}\right)^{2\Delta_{-}-3}+6\frac{\mathcal{E}(x)}{C_T}+ \mathcal{O}(r^{-1}) \nonumber 
\end{align}
for $\alpha \neq 0$, the surface integral is divergent for $1 < \Delta_- < 3/2$ while it is perfectly finite for $\Delta_{-}=1$ or $\Delta_{-}=0$.
In the case where the divergent term vanishes, the conformal boundary contribution of the surface integral is simply
\begin{equation}
\int_{\partial M_{\infty}}dA^{i}\partial_{i}P  = \frac{6}{C_T}\int_{\Sigma} d^2x\sqrt{\bar{g}} \mathcal{E}(x)
\end{equation}


\subsection{Non-extremal horizon}  \label{nearhorizon}
At finite temperature, the volume integral of $D^2P$ will automatically give the contribution to the on-shell euclidean action\footnote{Entirely or partially contributed to the on-shell action depending on the boundary conditions of the scalar field.}, hence it can also bound the free energy. In this situation, we assume the spacetime having killing horizon $\mathcal{H}$ with surface gravity $\kappa = 2\pi T$, then locally near the horizon the metric can be  written in terms of normal radial coordinate $r$ and horizon spatial coordinate $x^{a}$ as 
\begin{equation}
ds^2 = -\kappa^2r^2Q(r,x)dt^2+dr^2+\textit{\textg}_{ab}^{(\mathcal{H})}(r,x)dx^{a}dx^{b}
\end{equation}
where $Q(r,x)$ and $\text{\textg}_{ab}^{(\mathcal{H})}(r,x)$ is smooth function in $r^2$ at $r=0$ where the horizon is located at,
\begin{align}
Q(r,x) &= 1-\frac{1}{6}\left(\mathcal{R}-\frac{1}{2}(\mathcal{D\varphi})^2 \right)r^2+\mathcal{O}(r^4) \nonumber \\
\textit{\textg}_{ab}^{(\mathcal{H})}(r,x) &= \textit{\textg}_{ab}(x) +\frac{1}{2}\left( \mathcal{R}_{ab}-\frac{1}{2}\mathcal{D}_{a}\varphi\mathcal{D}_{b}\varphi-\frac{1}{2}\textit{\textg}_{ab}V(\varphi) \right)r^2+\mathcal{O}(r^4) \nonumber \\
\phi(r,x) &= \varphi(x) -\frac{1}{4}\left( \mathcal{D}^2\varphi-V'(\varphi)\right) r^2+\mathcal{O}(r^4)
\end{align}
Here $\mathcal{R}_{ab}$, $\mathcal{R}$ and $\mathcal{D}_a$ are the Ricci tensor, Ricci scalar and covariant derivative with respect to metric on the horizon $\textit{\textg}_{ab}(x)$.
With the above near horizon behaviour, $Z(x^k)$ and $g_{ij}(x^k)$ of optical geometry can be written as
\begin{align} \label{NRexpansion}
Z(r,x) &= \frac{\ell}{\kappa r}\left( 1+\frac{1}{12}\left(\mathcal{R}-\frac{1}{2}(\mathcal{D}\varphi)^2 \right)r^2+\mathcal{O}(r^4) \right)  \\
g_{rr}(r,x) &= \frac{1}{\kappa^2 r^2}\left( 1+\frac{1}{6}\left(\mathcal{R}-\frac{1}{2}(\mathcal{D}\varphi)^2 \right)r^2+\mathcal{O}(r^4) \right)  \\
g_{ab}(r,x) &= \frac{Z^2}{\ell^2}\textit{\textg}_{ab}^{(\mathcal{H})}(r,x) \nonumber \\
		&= \frac{1}{\kappa^2r^2}\left( \textit{\textg}_{ab}+\frac{1}{2}\left(\mathcal{S}_{ab}+\frac{1}{3}\textit{\textg}_{ab}\mathcal{S}-\frac{1}{2}\textit{\textg}_{ab}V(\varphi) \right)r^2+\mathcal{O}(\varphi^4) \right)
\end{align}
where 
\begin{equation}
\mathcal{S}_{ab}\equiv \mathcal{R}_{ab}-\frac{1}{2}\mathcal{D}_a\varphi\mathcal{D}_b\varphi \quad,\quad \mathcal{S}=\textit{\textg}^{ab}\mathcal{S}_{ab}
\end{equation}
Hence,by using (\ref{Peqn}) the near horizon expansion for $P$ is
\begin{equation}
P=-6\kappa^2+6\kappa^2\left( \frac{1}{\ell^2}+\frac{1}{2}\left(\mathcal{R}-\frac{1}{2}(\mathcal{D}\varphi)^2\right)\right)r^2+\mathcal{O}(r^4)
\end{equation}
where the surface gravity is $\kappa=2\pi T$. The near horizon contribution for integral of $D^2P$ over the whole optical geometry is 
\begin{align}
\int_{\mathcal{H}}dA^i\partial_iP &= \int_{\mathcal{H}}d^2x \sqrt{g^{(2)}} n^r(\partial_rP)\bigg{|}_{r=0} \nonumber \\
			&=  -\int_{\mathcal{H}}d^2x\frac{\sqrt{\textit{\textg}}}{\kappa^2r^2}\kappa r(1+\mathcal{O}(r^2))\left( 6\kappa^2\left( \frac{1}{\ell^2}+\frac{\mathcal{S}}{2}\right)2r+\mathcal{O}(r^3)\right) \bigg{|}_{r=0} \nonumber \\
			&= -12\kappa \int_{\mathcal{H}}d^2x\sqrt{\textit{\textg}}\left( \frac{1}{\ell^2}+\frac{1}{2}\left(\mathcal{R}-\frac{1}{2}(\mathcal{D}\varphi)^2\right) \right) \nonumber \\
			&= -24\pi T\left( \frac{A_\mathcal{H}}{\ell^2}+2\pi\chi(\mathcal{H})-\frac{1}{4}\int_{\mathcal{H}}(\mathcal{D}\varphi)^2 \right)
\end{align}
where 
\begin{equation}
A_\mathcal{H} = \int_{\mathcal{H}}d^2x\sqrt{\textit{\textg}} \quad,\quad \chi(\mathcal{H}) =\frac{1}{4\pi}\int_{\mathcal{H}}d^2x\sqrt{\textit{\textg}}\phantom{.}\mathcal{R}(\textit{\textg})
\end{equation}
are the area and Euler characteristic of $\mathcal{H}$ respectively, and 
\begin{align}
g^{(2)} &= \text{det}(g_{ab}) = \frac{\text{det}(\textit{\textg}_{ab})}{\kappa^4r^4}(1+\mathcal{O}(r^2)) \quad,\quad n^r=\frac{-1}{\sqrt{g_{rr}}} =-\kappa r(1+\mathcal{O}(r^2))
\end{align}


\subsection{Extremal horizon}   \label{nearextremalhorizon}
For pure gravity in the bulk, in the case that conformal boundary having null infinity and being at zero temperature(vanishing surface gravity) the IR geometry is extremal horizon whose metric near the horizon take the form \cite{Hickling:2014dra}\cite{Kunduri:2013ana}
\begin{equation}
ds^2 = -U(r,y)r^2dt^2+V(r,y)\left( \frac{dr}{r}+r\omega_a(r,y) dy^a \right)^2+h_{ab}(r,y)dy^ady^b  \label{extremal}
\end{equation}
where every function in the metric $U(r,y),V(r,y),\omega_a(r,y) , h_{ab}(r,y)$ are smooth function in terms of coordinate $(r,y)$ at $r=0$ where extremal horizon locates at and $y^a$ is coordinate on extremal horizon $\mathcal{H}_{\text{extr}}$.
\begin{align}
U(r,y) &= U^{(0)}(y)+ r U^{(1)}(y)+\mathcal{O}(r^2)  \nonumber \\
V(r,y) &= V^{(0)}(y)+ r V^{(1)}(y)+\mathcal{O}(r^2) \nonumber \\
\omega_a(r,y) &= \omega^{(0)}_{a}(y)+ r \omega^{(1)}_{a}(y)+\mathcal{O}(r^2) \nonumber \\
h_{ab}(r,y) &= h^{(0)}_{ab}(y)+ r h^{(1)}_{ab}(y)+\mathcal{O}(r^2) \nonumber 
\end{align}
Upon taking near horizon limit by $(t,r)\to (t/\epsilon, \epsilon r)$ then taking $\epsilon \to 0$ the near horizon geometry looks like
\begin{equation}
ds^2_{NH}= \psi(y)^2\left( -r^2dt^2+\frac{dr^2}{r^2} \right)+h^{(0)}_{ab}(y)dy^ady^b
\end{equation}
where $\psi(y)^2=U^{(0)}(y)=V^{(0)}(y)$ and $h^{(0)}_{ab}(y)$ are near horizon data of the metric.
Using the above metric for extremal horizon geometry (\ref{extremal}) we can deduce that function $Z(x^i)$ and optical metric $g_{ij}(x^i)$ in terms of functions in metric (\ref{extremal}) take the following form
\begin{equation}
Z(r,y^a) = \frac{\ell}{r\sqrt{U(r,y^a)}}
\end{equation}
\begin{equation}
g_{ij}(r,y^a) =\frac{V}{r^4U}\left( \begin{array}{cc}
1 & r^2\omega_b \\
r^2\omega_a & \frac{r^2}{V}h_{ab}+r^4\omega_a\omega_b
\end{array} \right) ,\quad 
g^{ij}(r,y^a) =\frac{r^4U}{V}\left( \begin{array}{cc}
1+r^2V\omega^2 & -V\omega^b \\
-V\omega^a & \frac{V}{r^2}h^{ab}
\end{array} \right)
\end{equation}
where $\omega^2 = h^{ab}\omega_a\omega_b$. Let's consider quantity $P$, thank to (\ref{Peqn}) we can translate $P$ in terms of optical geometry to extremal horizon geometry quantities using 
\begin{align}
\partial_iZ(r,y^a) &= (\partial_rZ,\partial_aZ)  \nonumber \\
		&= -\frac{Z}{2}\left( \frac{1}{r^2U}\partial_r\left(r^2U\right),\frac{\partial_aU}{U} \right)
\end{align}
together the above data for $Z(r,y^a)$ and $g^{ij}(r,y^a)$ 
\begin{equation}
P=6r^2U\Bigg{[}1-\frac{r^2}{4V}\left( \left( 1+r^2V\omega^2\right)\left( \frac{1}{r^2U}\partial_r\left( r^2U\right)\right)^2-2V\left( \frac{1}{r^2U}\partial_r\left( r^2U\right)\right)\omega^a\frac{\partial_aU}{U}+\frac{V}{r^2U^2}h^{ab}\partial_aU\partial_bU  \right) \Bigg{]}  \nonumber 
\end{equation} 
by expanding $P$ with near horizon expansion for $U,V,\omega_a,h_{ab}$ and the fact that $U_0(y)=V_0(y)=\psi(y)^2$. Near horizon expansion for $P$ is
\begin{equation}
P(r,y^a) = -6\left(1+h^{ab}_{(0)}\partial_a\psi\partial_b\psi-\psi^2 \right)r^2+\mathcal{O}(r^3)
\end{equation}
By which, its near extremal horizon contribution to the surface integral vanishes since 
\begin{align} 
\int_{\mathcal{H}_{\text{extr}}}dA^i\partial_iP &= \int_{\mathcal{H}_{\text{extr}}}d^2y\sqrt{h^{(0)}}n^{r}(\partial_rP)|_{r=0} \nonumber \\
			&= \int_{\mathcal{H}_{\text{extr}}}d^2y\sqrt{h^{(0)}}r^2(\partial_rP)|_{r=0} \nonumber \\
			&= 0
\end{align}
where $n^{r}=1/\sqrt{g_{rr}}=r^2(U/V)^{1/2}$ is radial component of unit spacelike normal vector. 

\section{On-shell action from optical frame variables}\label{Sonshell}

Using the trace of the equations of motion (\ref{main}) and (\ref{Veqn}) to substitute in the action then the bulk part of the on-shell action $S_{\text{bulk}}=S_{\text{grav}}+S_{\phi}$ becomes
\begin{align}
2\kappa_4^2 S_{\text{bulk}} &= -\int d^4x\sqrt{-g}V(\phi)  \nonumber  \\
		&= \int d^3x \int_{R_{IR}}^{R_{UV}} dr\frac{\ell^4}{Z^4}\sqrt{g_{(\text{op})}}\frac{Z^4}{\ell^2}D^2\left(\frac{1}{Z^2}\right)  \nonumber \\
	 \text{where $\sqrt{g_{(\text{op})}}$ is the square}&\text{ root determinant of the 3d optical metric $g_{ij}$.}\nonumber \\
		&= \ell^2\int d^3x \int_{R_{IR}}^{R_{UV}} dr\sqrt{g_{(\text{op})}}D^2\left(\frac{1}{Z^2}\right)  \nonumber \\
		&= \ell^2\int dt \bigg{[} \int d^2x \sqrt{g_{(2)}} n^r\partial_r\left(\frac{1}{Z^2}\right) \bigg{]}^{R_{UV}}_{R_{IR}} \quad ; \quad g_{(2)}=\text{det}(g_{ab}) \nonumber \\
		&= 2\kappa_4^2 S_{\text{bulk}}^{\text{(UV)}}-2\kappa_4^2 S_{\text{bulk}}^{\text{(IR)}}     \label{Sonshell}
\end{align}
where we assume that the optical metric takes the form $g_{ij}dx^idx^j=g_{rr}dr^2+g_{ab}dx^adx^b$. Near the boundary, we translate the FG expansion in appendix(\ref{nearboundary}) to the optical frame variables as the following
\begin{equation}
\begin{aligned}
n^r &= \frac{1}{\sqrt{g_{rr}}} = \frac{r^2}{\ell^2}\left( 1-\g_{(2)tt}\frac{\ell^4}{2r^2}-\g_{(2\Delta_{-})tt}\frac{\ell^{4\Delta_{-}}}{2r^{2\Delta_{-}}}-\g_{(3)tt}\frac{\ell^6}{2r^3}+\ldots  \right)  \\
\sqrt{g_{(2)}} &= \sqrt{\bar{g}}\bigg{(}1+(\g_{(2)}+3\g_{(2)tt})\frac{\ell^4}{2r^2}+ (\g_{(2\Delta_{-})ab}+\g_{(0)ab}\g_{(2\Delta_{-})tt})\frac{\ell^{4\Delta_{-}}}{r^{2\Delta_{-}}} \\
		&\phantom{====} +(\g_{(3)}+ 3\g_{(3)tt})\frac{\ell^6}{2r^3}+\ldots \bigg{)}   \\
\partial_r\left(\frac{1}{Z^2}\right) &= \partial_r\left( \frac{r^2}{\ell^4}-\g_{(2)tt} -\g_{(2\Delta_{-})tt}\left( \frac{\ell^2}{r} \right)^{2(\Delta_{-}-1)}-\g_{(3)tt}\frac{\ell^2}{r}+\ldots \right)  \\
&= \frac{2r}{\ell^2}+2(\Delta_{-}-1)\g_{(2\Delta_{-})tt}\frac{\ell^{4\Delta_{-}-4}}{r^{2\Delta_{-}-1}}+\g_{(3)tt}\frac{\ell^2}{r^2}+\ldots 
\end{aligned}
\end{equation}
where $\bar{g} = \text{det}(\bar{g}_{ab})$. Putting all these together, the UV part of the bulk on-shell action is
\begin{align}
2\kappa_4^2S_{\text{bulk}}^{\text{(UV)}} &= \int dt \bigg{[} \int d^2x\sqrt{\bar{g}}\left( \frac{2R_{UV}^3}{\ell^4} -\frac{3}{4}R(\bar{g})R_{UV}+\left( \frac{\Delta_{-}}{4}-\frac{3}{8} \right)\alpha^2\frac{\ell^{4\Delta_{-}-4}}{R_{UV}^{2\Delta_{-}-3}}+(3\g_{(3)tt}+\g_{(3)})\ell^2 \right)  \bigg{]}    \nonumber \\
 &\phantom{=} +\mathcal{O}\left(1/R_{UV}\right) \label{SbulkUV}  
\end{align} 
Next, we consider the boundary part of the action
\begin{equation}
\kappa_4^2S_{\text{bndy}} = \kappa_4^2(S_{GHY}+S_{\text{grav,ct}}+S_{\phi,\text{ct}}) 
\end{equation}
Since $m^2\ell^2=-2$ lies in the case of the scalar field with $m_{BF}^2\ell^2 < m^2\ell^2 \leq m_{BF}^2\ell^2+1/4$ where the counterterm action for scalar field with regular quantisation and alternative quantisation are different.
\begin{align}
\kappa_4^2S_{\text{bndy, reg.quant}} &= \int_{r=R_{UV}} d^3x\sqrt{-\g}\left( -K(\g)+\frac{2}{\ell}+\frac{\ell}{2}R(\g)+\frac{\Delta_{-}}{4\ell}\phi^2 \right) \\
\kappa_4^2S_{\text{bndy, alt.quant}} &= \int_{r=R_{UV}} d^3x\sqrt{-\g}\left( -K(\g)+\frac{2}{\ell}+\frac{\ell}{2}R(\g)-\frac{\Delta_{-}}{4\ell}\phi^2-\frac{1}{2}\phi n^r\partial_r\phi \right)
\end{align}
To calculate $S_{\text{bndy}}$ we need the following quantities in FG gauge
\begin{align}
\sqrt{-\g} &= \frac{r^3}{\ell^3}\sqrt{\bar{g}}\left(1+\g_{(2)}\frac{\ell^4}{2r^2}+\g_{(2\Delta_{-})}\frac{\ell^{4\Delta_{-}}}{2r^{2\Delta_{-}}}+\g_{(3)}\frac{\ell^6}{2r^3}+\ldots \right) \\
n^r &= \frac{r}{\ell}
\end{align}
where $-\text{det}(\g_{(0)})=\text{det}(\bar{g}_{ab})$ because $\g_{(0)tt}=-1$ in the optical frame
\begin{align}
K(\g) &= \frac{r}{\ell\sqrt{-\g}}\partial_r\sqrt{-\g} \nonumber \\
		&= \frac{3}{\ell}-\g_{(2)}\frac{\ell^3}{r^2}-\Delta_{-}\g_{(2\Delta_{-})}\frac{\ell^{4\Delta_{-}-1}}{r^{2\Delta_{-}}}-\frac{3}{2}\g_{(3)}\frac{\ell^5}{r^3}+\ldots  \\
R(\g) &= \frac{\ell^2}{r^2}R(\g_{(0)})+\mathcal{O}\left(r^{-3}\right) \\
\phi^2 &= \alpha^2\frac{\ell^{4\Delta_{-}}}{r^{2\Delta_{-}}}+2\alpha\beta\frac{\ell^6}{r^3}+\beta^2\frac{\ell^{4\Delta_{+}}}{r^{2\Delta_{+}}}+\ldots
\end{align}
Putting all these quantities into $S_{\text{bndy}}$ for regular and alternative quantisation respectively
\begin{equation}
\begin{aligned}
\kappa_4^2S_{\text{bndy,reg.quant}} &= \int d^3x\sqrt{\bar{g}}\bigg{[} -\frac{R_{UV}^3}{\ell^4}+\frac{3}{8}R(\bar{g})R_{UV}+\left(-\frac{\Delta_{-}}{8}+\frac{3}{16}\right)\alpha^2\frac{\ell^{4\Delta_{-}-4}}{R_{UV}^{2\Delta_{-}-3}} \\
					&\phantom{========}+\left(\frac{1}{2}\Delta_{-}\alpha\beta\ell^2+\g_{(3)}\ell^2\right) \bigg{]}+ \mathcal{O}\left(1/R_{UV}\right) \\
\kappa_4^2S_{\text{bndy,alt.quant}} &= \int d^3x\sqrt{\bar{g}}\bigg{[} -\frac{R_{UV}^3}{\ell^4}+\frac{3}{8}R(\bar{g})R_{UV}+\left(-\frac{\Delta_{-}}{8}+\frac{3}{16}\right)\alpha^2\frac{\ell^{4\Delta_{-}-4}}{R_{UV}^{2\Delta_{-}-3}} \\
					&\phantom{========}+\left(\frac{1}{2}\Delta_{+}\alpha\beta\ell^2+\g_{(3)}\ell^2\right) \bigg{]}+ \mathcal{O}\left(1/R_{UV}\right)
\end{aligned}
\label{Sbndy}
\end{equation}
At finite temperature, the IR geometry ends on a killing horizon with surface gravity $\kappa=2\pi T$ and the on-shell action will also get a contribution from killing horizon(s) as well, using the near horizon expansion in (\ref{NRexpansion})  the IR part of $S_{\text{bulk}}$ in (\ref{Sonshell}) is\footnote{With normal vector pointing outward such that $n^r = \kappa r(1+\mathcal{O}(r^2))$ instead, because we write the integral as $S_{bulk} \sim S_{UV}-S_{IR} $.}

\begin{align}
2\kappa_4^2S_{\text{bulk}}^{(\text{IR})} &= \int dt\int_{r=R_{IR}}d^2x\sqrt{g_{(2)}}n^r\partial_r\left(\frac{1}{Z^2}\right) \nonumber \\
			&= \int dt \int_{r=0} d^2x \frac{\sqrt{\textit{\textg}}}{\kappa^2r^2}\kappa r \frac{2\kappa^2r}{\ell^2}(1+\mathcal{O}(r^2)) \nonumber \\
			&= \int dt \int_{r=0} d^2x \sqrt{\textit{\textg}} \frac{2\kappa}{\ell^2}  \nonumber \\
			&= \int dt \frac{4\pi T}{\ell^2}A_{\mathcal{H}}  \label{SbulkIR}
\end{align}
where $A_{\mathcal{H}}=\int d^2x\sqrt{\textit{\textg}}$ is the horizon area.

Having obtained $S_{\text{bulk}}^{(\text{UV})}$, $S_{\text{bndy}}$ and $S_{\text{bulk}}^{(\text{IR})}$ in (\ref{SbulkUV}), (\ref{Sbndy}) and (\ref{SbulkIR}) respectively, the renormalised on-shell action $S^{(\text{ren})}_{\text{on-shell}}$ is then
\begin{align}
S^{(\text{ren})}_{\text{on-shell}} &= S_{\text{bulk}}^{(\text{UV})}-S_{\text{bulk}}^{(\text{IR})}+S_{\text{bndy}} \nonumber \\
			&= \frac{(\ref{SbulkUV})}{2\kappa_4^2} - \frac{(\ref{SbulkIR})}{2\kappa_4^2}+\frac{(\ref{Sbndy})}{\kappa_4^2} \nonumber \\
			&= \int dt \bigg{[} \frac{\ell^2}{16\pi G_4}\left(\int d^2x\sqrt{\bar{g}}\left( 3\g_{(3)tt}+3\g_{(3)}+\Delta_{\mp}\alpha\beta \right)\right)-TS \bigg{]}
\end{align}
where the coefficient of $\alpha\beta$ term is $\Delta_{-}$ or $\Delta_{+}$ for the scalar field subjected to regular or alternative quantisation respectively and the entropy is $S=\frac{A_{\mathcal{H}}}{4G_4}$. Then we can make a transition to the euclidean signature by making the substitution
\begin{align}
t \to i\tau \quad &,\quad iS_{L} \to -S_{E} \nonumber \\
\text{integration domain } &:\quad 0 \leq \tau \leq \beta'
\end{align}
where $\beta'=1/T$, then the free energy is
\begin{align}
\beta' F &= S_{\text{E,on-shell}}^{(\text{ren})} \nonumber \\
F &= \frac{\ell^2}{16\pi G_4}\left(\int d^2x\sqrt{\bar{g}}\left( 3\g_{(3)tt}+3\g_{(3)}+\Delta_{\mp}\alpha\beta \right)\right)-TS 
\end{align}

\paragraph{Relevant scalar} For the scalar field with mass $m^2\ell^2=-2$, the free energy for different quantisation schemes are as follows:
\begin{enumerate}
\item Regular quantisation(Dirichlet BC)
\begin{equation}
F_{\text{Dirichlet}}=\frac{\ell^2}{16\pi G_4}\left(\int d^2x\sqrt{\bar{g}}\left( 3\g_{(3)tt}+3\g_{(3)}+\alpha\beta \right)\right)-TS 
\end{equation}
\item Alternative quantisation(Neumann BC)
\begin{equation}
F_{\text{Neumann}}=\frac{\ell^2}{16\pi G_4}\left(\int d^2x\sqrt{\bar{g}}\left( 3\g_{(3)tt}+3\g_{(3)}+2\alpha\beta \right)\right)-TS 
\end{equation}
\end{enumerate}
where, for both cases, the trace of $\g_{(3)\mu\nu}$ is $\g_{(3)}=-\frac{2}{3}\alpha\beta$.

\paragraph{Massless scalar}On the other hand, for the scalar field which corresponds to the marginal deformation $\ell^2V(\phi)=-6+\mathcal{O}(\phi^3)$, the calculation can be repeated in the same fashion as the massive scalar field case but with the change of $S_{\phi,\text{ct}}$ to be 
\begin{equation}
S_{\phi,\text{ct}} = -\frac{1}{4\kappa_4^2}\int_{\partial\mathcal{M}}d^3x\sqrt{-\g}\g^{\mu\nu}\partial_{\mu}\phi\partial_{\nu}\phi
\end{equation}
In this case, the bulk part of the onshell action in the IR takes the same form as (\ref{SbulkIR}) but the $S_{\text{bulk}}^{(IR)}$ takes the following form
\begin{align}
2\kappa_4^2S_{\text{bulk}}^{\text{(UV)}} &= \int dt \bigg{[} \int d^2x\sqrt{\bar{g}}\left( \frac{2R_{UV}^3}{\ell^4}+\left( -\frac{3}{4}R(\bar{g})+\frac{3}{8}(\partial\alpha)^2 \right)R_{UV}+(3\g_{(3)tt}+\g_{(3)})\ell^2 \right)  \bigg{]}    \nonumber \\
 &\phantom{=} +\mathcal{O}\left(1/R_{UV}\right) \label{SbulkUVm0}  
\end{align} 
The boundary contribution of the on-shell action is 
\begin{equation}
\kappa_4^2S_{\text{bndy,}m^2=0} = \int d^3x\sqrt{\bar{g}}\bigg{[} -\frac{R_{UV}^3}{\ell^4}+\left(\frac{3}{8}R(\bar{g})-\frac{3}{16}(\partial\alpha)^2\right)R_{UV}+\g_{(3)}\ell^2\bigg{]}+ \mathcal{O}\left(1/R_{UV}\right) \label{Sbndym0}
\end{equation}
then the on-shell action can be obtained from 
\begin{align}
S^{(\text{ren})}_{\text{on-shell},m^2=0} &= S_{\text{bulk}}^{(\text{UV})}-S_{\text{bulk}}^{(\text{IR})}+S_{\text{bndy}} \nonumber \\
			&= \frac{(\ref{SbulkUVm0})}{2\kappa_4^2} - \frac{(\ref{SbulkIR})}{2\kappa_4^2}+\frac{(\ref{Sbndym0})}{\kappa_4^2} \nonumber \\
			&= \int dt \bigg{[} \frac{\ell^2}{16\pi G_4}\left(\int d^2x\sqrt{\bar{g}}(3\g_{(3)tt})\right)-TS \bigg{]}
\end{align}
since $\g_{(3)}=0$ for the solution with massless scalar. Then the free energy can be obtained in the same fashion as the previous case
\begin{equation}
F_{m^2=0} = \frac{\ell^2}{16\pi G_4}\left(\int d^2x\sqrt{\bar{g}}( 3\g_{(3)tt} )\right)-TS 
\end{equation}


\bibliography{scalarbib}{}

\bibliographystyle{JHEP}

\end{document}